 \documentclass[final,11pt,3p,times]{elsarticle}
\usepackage{graphicx}
\usepackage{amssymb}
\usepackage{lineno}
\usepackage{amsmath,graphicx}
\usepackage{amsfonts}
\usepackage{color}
\usepackage{amssymb}
\usepackage{amsthm}
\usepackage{cite}
\usepackage{array}

\usepackage{dsfont}
\usepackage{algorithm, algorithmic}
\usepackage{bm}

\newcommand{\revis}{\textcolor{black}}
\newcommand{\newrev}{\textcolor{black}}

\newcommand{\SPrevis}{\textcolor{black}}
\newcommand{\SPrevisNew}{\textcolor{black}}

\journal{Signal Processing}

\begin{document}

\begin{frontmatter}


%% \title{A Distributed Particle-PHD Filter with Arithmetic-Average PHD Fusion\vspace{2mm}}
\title{\LARGE From Target Tracking to Targeting Track: A Data-Driven Yet Analytical %, Least Squares %Data-Driven %Bare-handed %, Yet Heuristic
Approach to Joint Target Detection and Tracking}
%% \thanks{This work was partially supported by the Marie Sk\l{}odowska-Curie Individual Fellowship (H2020-MSCA-IF-2015) under Grant 709267 and by the Austrian Science Fund (FWF)
%% under Grants P27370-N30 and P32055-N31.}}


 \author[label1]{\large Tiancheng Li}  % ,label2
 \author[label1]{\large Yan Song}  % ,label2
 \author[label3]{\large Hongqi Fan}
 \address[label1]{Key Laboratory of Information Fusion Technology (Ministry of Education), School of Automation, Northwestern Polytechnical\\University, Xi'an 710072, China.
 E-mail: t.c.li@nwpu.edu.cn.} %, zheng.hoo@mail.nwpu.edu.cn, xyue@mail.nwpu.edu.cn \vspace{1.5mm}}
%\address[label2]{AIR Institute, Salamanca 37188, Spain\vspace{1.5mm}}
\address[label3]{National Key Laboratory of Science and Technology on ATR, National University of Defense Technology, Changsha 410073, Hunan, China. E-mail: fanhongqi@nudt.edu.cn \vspace{0mm}}
%% \address[label4]{Brno University of Technology, 60190 Brno, Czech Republic}

\begin{abstract}
\renewcommand{\baselinestretch}{1.17}\small
This paper addresses %proposes a data-driven approach to
 the problem of real-time
detection and tracking of a non-cooperative target in the challenging scenario %with
with almost no a-priori information about target birth, death, dynamics and detection probability. %, and clutter rate. %, namely target tracking in the dark (TTinDark). The ``dark'' is twofold regarding both model and data.
%That is, the target may appear, move, disappear and re-appear randomly in both space and time domains of unknown statistical information. %Nor is there any a-priori target kinematic model information.
Furthermore, there are false and missing data at an unknown yet low rate in the measurements. %, namely the target may be undetected (generating no measurement) and clutter-measurements are randomly generated over the scenario at each scan, for both of which there is no a-priori statistical information too.
The only information given in advance is about the target-measurement model and the constraint that there is no more than one target in the scenario. % which has a low initial velocity.
To solve these challenges, we model the movement of the target by using a polynomial trajectory function of time (T-FoT)\SPrevis{, which aims to estimate the \textit{continuous-time trajectory} of the target rather than a series of discrete-time point estimates as is done in most existing filters/trackers.} %online estimated by using the measurements in time series  % and thereby reformulates the conventional smoothing, filtering and tracking jointly as an online curve/FoT fitting problem.
% Instead of estimating the \textit{discrete-time state} of the target based on a sophistically designed SSM% which deals with ``tracking'' as a standard state estimation problem
%, we are actually more interested in estimating the \textit{continuous-time trajectory/track} of the target in the context of target tracking. \newrev{That is, we seek direct estimating the trajectory.
Data-driven T-FoT initiation and termination strategies are proposed for identifying the (re-)appearance and disappearance of the target. During the existence of the target, real target measurements are distinguished from clutter if the target indeed exists and is detected, %and misdetection is declared otherwise,
in order to update the T-FoT at each scan for which we design a least-squares estimator. \newrev{Overall, our approach is Markov-free, data-driven yet analytical.} % to be able to resolve the ``dark'' situation:. %, although the parameters needed are determined in an more or less heuristic manner.
Simulations %based on 1000 ground truths for
using either linear or nonlinear systems are conducted to demonstrate the effectiveness of our approach in comparison with the Bayes optimal Bernoulli filters. The results show that our approach is comparable to the perfectly-modeled filters, even outperforms them in some cases while requiring much less a-priori information and computing much faster.
\vspace{1mm}
\end{abstract}


\begin{keyword}
\renewcommand{\baselinestretch}{1.17}\small
Target tracking, target detection, trajectory fitting, least-squares estimation, Bayes inference
%% , partial consensus.

\end{keyword}

\end{frontmatter}


\vspace{-1mm}

\section{Introduction}
Target
tracking that involves the online estimation of the trajectory of a target has been a long standing research topic and plays a significant role in aerospace, traffic, defence, robotics, etc. \citep{Bar-Shalom01} %
\SPrevis{In this work, we focus on an important class of targets with simple and smooth {trajectory} such as airplane, train, ship and so on. Here, the smoothness of the trajectory is closely related with the dynamical equations of the target based on differential calculation.}
The standard approach since the ground-breaking Kalman filter (KF) \citep{kalman60}, is to design a state space model (SSM) consisting of a Markov-jump model to describe the dynamics of the target and a measurement model to relate the measurement \SPrevis{of the sensors} with the state of the target. In addition, models are needed to characterize the \SPrevis{background} clutter-measurements and misdetection events. However, in many cases \SPrevis{especially those for} non-cooperative targets \citep{Blacknell13}, these models except the measurement function are unavailable in advance and intractable to be identified accurately online. \SPrevis{This leads to a great challenge to the use of any filters.} %, for non-cooperative targets, there is a great challenge to identify the target dynamics in order to properly set up a filter.

Instead of estimating the \textit{discrete-time state} of the target based on a sophistically designed SSM% which deals with ``tracking'' as a standard state estimation problem
, we are actually more interested in estimating the \textit{continuous-time trajectory/track} of the target in the context of target tracking. \newrev{That is, we seek direct estimating the trajectory. % As to be addressed in this paper,
Once such a trajectory is obtained, the position (and also the velocity and acceleration which correspond to the first and second order derivatives of the position against time)} %(and its first and second order derivatives over time which indicate the velocity and acceleration, respectively) %; see, e.g., \citep{Pilte17}
of the target at a particular time can be easily inferred from the trajectory. \SPrevis{Moreover, high-level important information such as the class of the target, the motion model/pattern/feature and so on may be able to be inferred from the continuous-time trajectory but not from a series of discrete point estimates. Existing approaches} to target trajectory estimation can be categorized in the following two groups, depending on modeling the trajectory whether as a discrete-time series of position points or as a continuous-time curve. \newrev{The latter which describes the movement of the target in continuous time is the focus of this paper.}

\subsection{Discrete-time Trajectory Estimation}
\SPrevis{There are various SSM-based studies that recursively estimate the discrete-time-series state set %(namely discrete-time trajectory)
based on the measurement sequence. %\citep{Stemler09, Smith10,}
%%In the single target case,
Typically, it is given by a trajectory of a prescribed dynamical model such that the output of the model best fits a series of measurements. This is often referred to generally as data assimilation which have found many applications \citep{Wang13, Rosenthal17,Carrassi18}.} %Talagrand87,Brocker12,  % Dimet86,
%searches for the maximum of the posterior density function by assuming certain SSMs.
%%For linear systems, discrete-time trajectory estimation has direct connection to %Gauss' Least-squares (LS) estimate and
%%Kolmogorov-Wiener's interpolation and extrapolation of a sequence \citep{Singpurwalla18}. %\citep{Plackett50}
Different from the stochastic modeling of the state process, %Judd etc. presented a series of non-sequential/optimization-based estimation and forecasting works in the context of chaotic systems, e.g., \citep{Judd00, Judd08, Judd09forecasting, Judd15, Smith10}, which %, as a core difference to the stochastic approach,
% remove the use of the state transition noise. Actually, similar
deterministic Markov-jump models, namely using no random variable $\mathbf{w}_k$ in \eqref{eq:Markov}, have been used in the so-called shadowing filters/smoothers \citep{Judd09failure,Zaitouny19},  %Stemler09, Judd00,Judd08,Judd15,noise reduction methods \citep{Kostelich93},
moving horizon estimator \citep{Michalska95,Rao03} and Gauss-Newton filter \citep{Morrison2012tracking, Nadjiasngar13}.
In particular, the Gauss-Newton filter that models the state transition by a deterministic differential equation is Cram\'{e}r-Rao consistent (providing minimum variance) \citep{Morrison2012tracking}.
These approaches, mainly designed for some specific systems, avoid the difficulties to accurately model the process noises and % \citep{Dunik17,Zhang20} %Interestingly, Judd's shadowing filter yields more reliable and even more accurate performance than the Bayesian filters - \textit{however, a fairer comparison should be made between shadowing filters with Bayesian smoothers, using the same amount of measurement data} - in the case when the nonlinearity is significant, but the noise is largely measuremental \citep{Judd09failure}, or when the objects do not typically display any significant random motions at the length and the time scales of interest \citep{Judd15}.
%In particular, the Gauss-Newton filter that models the state transition by a deterministic differential equation is Cram\'{e}r-Rao consistent (providing minimum variance)\citep{Morrison2012tracking}.
%Despite their Markov assumptions, these approaches %, akin to our fitting approach,
are advantageous in handling constraints and measurement singularities. They, however, still rely heavily on the Markov-jump model, which is essentially vulnerable to target maneuvering and unknown input (e.g., non-zero acceleration in the context of target tracking). To handle unknown inputs/noises in the motion of the target, an alternative is the minimum model error estimator \citep{Mook88,Crassidis97} which, different from the classic KF, requires no a-priori statistics on the form of the model error but determines it as a part of the solution. %Obviously, when the process noise is significant, these approaches ignoring it %We will not pay much attention to works in this direction.% than a stochastic recursive filter.

\revis{
In the multi-target case, by adding an unique label to each target, the target trajectory can be consequently given by the time-series labelled states \citep{Delong12,Vo19multiscan}.
%This leads to results similar in spirit to the classic data-association based tracking approaches such as the well-known %(probabilistic)
%multiple hypothesis tracker. %\citep{Vo15mtt}. %,Vo19multiscan Garcia-Fernandez14,
Despite its promising performance, an insightful discussion on the optimality of linking discrete-time points is given in \citep{Chen18}. Another relevant approach models the trajectories of a random number of targets as a random finite set (RFS) to be estimated \citep{Garcia-Fernandez19}. %, and iteratively minimizing a cost function which is the negative of the logarithm of the posterior density function.
Compared to the point-state RFS, the trajectory RFS requires much higher computation and needs to solve the problem of trajectories of different lengths in the same RFS.}

\subsection{Continuous-time Trajectory Fitting}
Continuous-time trajectory curve described by a function contains more information than the discrete-time point set.
For example, one can learn more feature or class information of the target from the continuous-time trajectory rather than from the discrete-time point set. %and is therefore more useful. %and is therefore more useful. %estimation via data fitting has been studied in different disciplines, which has the inherent advantages of dealing with outliers or missed detections \citep{Fan03}, especially when adequate analytical solutions may not exist.
\SPrevis{In fact, signal processing stems from the interpolation and extrapolation of a sequence of data that were viewed as a realization of a stochastic process \citep{Singpurwalla18}. }% Data fitting is a self-contained mathematical problem and a prosperous research theme by its own, which has proven to be a powerful and universal method for pattern learning and time series data prediction, especially when adequate analytical solutions may not exist.
%While there have been considerable relevant works, e.g., sum-of-norms optimization \citep{Rao03,Ohlsson12,Carrillo15,Yang17} %Ohlsson10,
%based on the SSM. %What is also notable is a family of recursive least square (RLS) estimators \citep{Chan04,Vahidi05,Bhotto11}, which reformulated in the state-space form was recognized a special case of the KF\citep{Sorenson70,Sayed94}.% %Earlier retrospective review of the LS estimation and its connection to the KF can be found in \citep{Sorenson70}.
% %that is the optimum linear tracking scheme based on second-order statistics
% Given guaranteed model constraint, the so-called optimal model error can be explicitly accounted for.
Data regression or fitting is a self-contained mathematical problem and a prosperous research theme by its own, which has proven to be a powerful and universal method for pattern learning and time series data prediction, \revis{and has the inherent advantages of dealing with outliers or missed detections \citep{Fan03},} especially when adequate analytical solutions may not exist. %, which has a much longer history than that of the KF.
In particular, there have been a large number of efforts devoted to trajectory curve fitting among different dimensions of the state space in various ways, e.g., \citep{El-Hawary95,Dimatteo01,Hadzagic11,Wang10,Liu14,Furgale15,Delong12,Milan16,Li16f4s}, %Bibby10, Lovegrove13
most of which, however, are \textit{non-recursive} over time and not designed for online tracking. %in the sense of Bayesian inference \citep{Hadzagic11,Dimatteo01} or externally to a recursive filter \citep{El-Hawary95,Wang10,Liu14,Li16f4s}. %In particular, the advantages of the MME estimator over the KF are that no a priori statistics on the form of the model error are required, which is determined as part of the solution, and that the states estimates are free of jump discontinuities while the approach is based on SSM\citep{Mook88,Crassidis97}. In \citep{Hadzagic11}, the trajectory is approximated by a cubic spline with an unknown number of knots in 2D position plane, and the function estimate is determined from positional measurements which are assumed to be received in batches at irregular time intervals. For the data drawn from an exponential family, the spline knot con-figurations (number and locations) are changed by reversible-jump Markov chain Monte Carlo \citep{Dimatteo01}. %, Dimatteo01
% maximum likelihood estimation (MLE) \citep{Anderson-Sprecher96}
%Continuous-time trajectory fitting has received attention in the context of mobile robot simultaneous localization and mapping \citep{Bibby10, Lovegrove13,Furgale15} and visual tracking \citep{Delong12, Milan16}. %In most of these approaches, the fitting is carried out purely in the state space , mostly in the planar position space,% where there are innumerable great achievements in dealing with various intractable tracking issues such as target occlusions and collisions.
Differently, as the few attempts that assume a spatio-temporal trajectory for tracking, \citep{Wang94,Anderson-Sprecher96,Wang_Zhu99,Liu_Zhu10,Zheng16} %Wang_Zhu98
designed fitting cardinal trajectory-splines for each dimension. % with the bearing data in the maximum likelihood estimation manner.
The resulting trajectory is a continuous-time {function of state}, the same to our approach \citep{Li19stf,Li18MoC,Li19FoTClutter}; see Section \ref{sec:background}.
Even more advanced, machine learning and neural networks were employed for trajectory fitting \citep{Hamed13,Thormann17,Yan19RNN,Pinto21}, %Rudy19RNNidentification
which, however, typically require an even larger amount of training data generated from proper a-priori models; \SPrevis{a cutting-edge review of Bayesian learning for data analysis can be found in \citep{Cheng22}.} \SPrevis{These approaches are usually not analytical and suffer from lack of interpretability of the parameters in the networks.}
Either way, existing trajectory fitting approaches %are mostly batch and not suitable for online tracking, not to say taking
have not taken into account realistic tracking issues such as clutter measurements, missed detection and target death, etc. \newrev{as we will do in this paper.}
%In the former, the robot motion is usually under the user's control (called proprioceptive sensor data) and the continuous-time trajectory representation makes it easy to deal with asynchronous measurements and constraints. In the latter, starting and/or ending points may be specified for the trajectory as constraint, as we have done in \eqref{eq:FoT_optimization}.
%When multiple targets exist, DA becomes necessary.
%The tracking problem is treated as the discrete-continuous optimization with label costs\citep{Milan16}, for which the key is generating all the trajectory hypotheses having a reasonable low label cost based on a variety of DA rules, for which the design of the label cost takes the critical issues such as the targets' dynamics, occlusions and collisions into account. However, only linear fitting is involved. % In \citep{Milan16}, the DA and trajectory estimation are formulated jointly in the minimization of a consistent discrete-continuous cost function, based on the multi-model fitting framework proposed earlier in \citep{Delong12}, where trajectories are modeled by piecewise polynomials.

What is more, it is necessary to identify in real-time whether the target is present or absent, especially when the target is non-cooperative. This task is often referred to separately as ``target detection" and is a prerequisite for tracking. As such, a reasonable solution is to address target detection and tracking jointly, \newrev{as required in the context of realistic target tracking. Following this thinking, the terminology of ``tracking'' means more than ``filtering'' does.} %So, this letter is more focused on the ``track/trajectory estimation''.
%There have been various works devoted to ``trajectory estimation'' and ``joint detection and tracking".

\subsection{Joint Detection and Tracking (JDT)}
\SPrevis{An indispensable task in realistic target tracking is target detection that involves identifying whether the target is present (or how many targets are present in the multi-target case) since the target may randomly appear in or disappear from the field of view.}
There have been a plurality of SSM-based JDT approaches, e.g., \citep{Musicki94IPDA,Tonissen96,Jiang14,Mallick15}.
%, which is also referred to as track-before-detect \citep{Buzzi05TBD,Mallick14book,Baser15} %Moyer11HoughTransform
%that works with raw measurements for track/trajectory extracting.
One of the most attractive theories for this purpose is the RFS \citep{Mahler07book} of which the most known single-target detection and tracking filter is the Bernoulli filter (BF) that is exact Bayes optimal \citep{Mahler07book,Vo12Bernoulli,Ristic13Bernoulli}. %,Li19Bernoulli
Given exact statistics about the target birth, death, dynamics and detection probability and clutter models and rate, such a single target tracking problem can be properly solved by the BF.
This, however, is often too ideal to be true for non-cooperative targets as these statistical information are generally unknown, time/space-varying and can hardly be exactly characterized. %For example, the target detection probability is %Despite fruitful effects devoted to online estimating parameters with the target state for filter robustness or adaptability, all filters have to be properly modelled regarding the targets, the sensor and the background. %While there are fruitful works addressing the lack of statistics about false alarm, missed detection, or random birth and death of the target, no works address all of these issues jointly in a single framework.
In contrast,
%According to our best knowledge, few work addresses the lack of
it becomes much challenging when there is little a-priori statistical information about the target %about target birth, death, dynamics and detection probability, and clutter rate simultaneously
which is just the actual need in many problems.
While many effects have been devoted to online estimating one or two of these issues, % for robustness or adaptability,
e.g., fruitful achievements for noise identification alone \citep{Dunik17,Zhang20}, no work addresses all in one. %, to name a few,Zhang19
%In the non-cooperative case that we consider here, there is almost no a-priori information available except for the measurement function.
This motivates our work in this paper.
%The goal of this paper is to develop a data-driven JDT method to fill this void. %They are the root causes for false alarm and missed detection and are unavoidable in practice.

%To fill this void, the T-FoT-oriented tracking approach is in this paper extended to accommodate random finite set measurements that are affected by false and missing data. That is, the number of measurements received at each scan is random and finite. What is more challenging, little is known about the statistics (e.g., ratios) for missed detection, false alarm, and even for the measurement noise, all of which are inherently time-varying. This prevents straightforward application of a conventional filter.

\subsection{Contribution and Paper Organization}
\SPrevis{In this paper, the target may randomly appear, disappear and re-appear anytime and anywhere in the surveillance scenario while the number of targets is no more than one. The only information available a-priori is the statistics of the measurements.} %We have to remind that, since there is totally statistical information available, it is unsurprising that our approach has to apply several
% and are particularly devoted to the design of an algorithm using the least.
For non-cooperative target tracking with little a-priori information, we earlier proposed to model the target movement by using a trajectory function of time (T-FoT) which reformulates the tracking as an online curve fitting problem \citep{Li19stf,Li18MoC} for which the least-squares (LS) approach plays a key role. Issues of false and missing data, \newrev{as well as the connection of the approach with the classic KF}, have been further addressed in \citep{Li19FoTClutter}, based on the prerequisite that the target birth/initial position, velocity and starting time are given in advance and that the target always exists in the scenario. %and verified the approach on a linear single-sensor system.
In this work, we relax these restrictive assumptions and further address the random (in both the state space and time domain) birth, death and even re-appearance of the target, improving the algorithm for better identifying false and missing measurements. \SPrevis{All aim at a computationally efficient, data-driven JDT approach that suits the challenging, unknown background.}

%To further take into account the issues of, the T-FoT-based tracking in clutter has recently been resolved in \citep{Li19FoTClutter}, in which we propose clutter removal and misdetection compensation procedures. But the work is greatly based on the prerequisite that the target has been detected and an initial track has been obtained. It is unable to deal with (re-)appearance and disappearance of the target. In the what follows, we address the random appearance and disappearance of the target, to integrate the proposed clutter removal and misdetection compensation procedures in order to get a

The remainder of the paper is organized as follows. Preliminary work is briefly addressed in Section \ref{sec:background}. Scenario assumptions and models we consider are given in Section \ref{sec:model}. The proposed approach is given in Section \ref{sec:proposal}.
Simulation results of our approach are given in Section \ref{sec:simulation}, in comparison with the state-of-the-art, properly modelled BF \citep{Vo12Bernoulli,Ristic13Bernoulli}. The paper is concluded in Section \ref{sec:conclusion}.

%\textcolor{blue}{[[In the following, materials that cannot be found in previous works are marked in blue print or noted in red print.]]}

\section{Preliminaries}\label{sec:background}
\subsection{Conservational Models and An Alternative}
In this work, we focus on the single target detection and tracking problem.
This problem is usually solved based on the SSM with additive noises which can be described as follows
\begin{align}
\mathbf{x}_k &= f_k(\mathbf{x}_{k-1}) + \mathbf{w}_k, \label{eq:Markov}\\
\mathbf{y}_k &= h_k(\mathbf{x}_k) + \mathbf{v}_k. \label{eq:MeasurementModel}
\end{align}
where $k\in \mathbb{N}$ indicates the time-instant, $\mathbf{x}_k\in \mathbb{R}^{D_\mathbf{x}}, \mathbf{y}_k\in \mathbb{R}^{D_\mathbf{y}}$, $\mathbf{w}_k \in \mathbb{R}^{D_\mathbf{x}}$ and $\mathbf{v}_k\in \mathbb{R}^{D_\mathbf{y}}$ denote the $D_\mathbf{x}$-dimensional state, the $D_\mathbf{y}$-dimensional measurement and their respective noises \citep{Dunik17,Zhang20} at time $k$, respectively.

The SSM facilitates recursive optimal/suboptimal filtering \citep{Sarkka13book,Li17AGC} in the Bayesian fashion whose performance, however, heavily relies on the match between the Markov model \eqref{eq:Markov} and the true target dynamics which is intractable to be identified \citep{Tobar15}, % typically time-varying (as called target maneuvering) and
as the terminology \textit{non-cooperative} indicates. From time to time, non-cooperative target may maneuvers \citep{Li03,Li05} or there is unknown input \citep{Li17AGC} %(namely changes its mode) %\citep{Li03}
and so on, adding difficulties to precisely capture its kinematic model as in \eqref{eq:Markov}. This leads to a fundamental challenge in practice.
%, for which
%the Bayesian posterior is given by
%\begin{equation} \label{eq:Bayes}
%% \mathbf{\hat{x}}_k = \int \mathbf{x} p(\mathbf{x}_k|\mathbf{y}_{1:k}) d\mathbf{x}
%p(\mathbf{x}_k|\mathbf{y}_{1:k}) = \frac{p(\mathbf{y}_k|\mathbf{x}_{k})p(\mathbf{x}_k|\mathbf{x}_{k-1},\mathbf{y}_{1:k-1})} {p(\mathbf{y}_k|\mathbf{y}_{1:k-1}) }, % \int p(\mathbf{y}_k|\mathbf{x}_{k})p(\mathbf{x}_k|\mathbf{x}_{k-1},\mathbf{y}_{1:k-1})d\mathbf{x}_k
%\end{equation}
%where, lying at the core of the Bayes inference, $p(\mathbf{x}_k|\mathbf{x}_{k-1},\mathbf{y}_{1:k-1})$ and $p(\mathbf{y}_k|\mathbf{x}_{k})$ are the prediction density and the likelihood density, %relying on the dynamic model \eqref{eq:Markov} and measurement model \eqref{eq:MeasurementModel},
%respectively.

%Rather than investigating any SSM-based filters (of which there have already been a plethora),
%we proposed to model the target movement by using a trajectory function of time (T-FoT) $\mathbf{x}_t= {f}(t)$ and thereby reformulates the conventional tracking as an online curve fitting problem \citep{Li19stf,Li18MoC}.
To eschew this challenge, we consider modeling the ground truth of the target motion by an engineering-friendly, spatio-temporal T-FoT $\mathbf{x}_t= {f}(t)$, where the target state $\mathbf{x}_t$ is limited to its coordinate position in the following unless otherwise stated. Moreover, the velocity and acceleration can be obtained from the derivatives of the T-FoT over time. %, which best fits the sensor data as in \eqref{eq:FoT_optimization}.
In particular, we parameter the T-FoT ${f}(t)$ by
${F}(t;\mathbf{C})$ using a set of coefficients $\mathbf{C}$ that determine the trajectory based on the available measurements. This may be expressed as follows
\begin{equation}\label{eq:ft_Ft}
  \mathbf{x}_t = {F}(t;\mathbf{C}) + \mathbf{\epsilon}(t) ,
\end{equation}
where $t\in\mathbb{R}^+$ indicates the (positive) continuous-time, $\mathbf{\epsilon}(t)$ denotes the T-FoT residual function-of-time, namely the approximate error of ${F}(t;\mathbf{C})$ to the real T-FoT ${f}(t)$.

% which at discrete-time instants takes the same value as $k$.

%We are particularly interested in the target that moves on a ``smooth" trajectory, such as aircraft, train, cruise ship, etc.

By using the T-FoT, what is estimated now is the trajectory parameters $\mathbf{C}$. % which is sufficiently general to define any types of trajectory rather than the discrete-time \textit{point-state} $\mathbf{x}_k$.
This formulation is naturally appealing to an important type of targets with smooth motion patterns such as aircraft, train, cruise ship, etc. %and eschews the intractability of modeling the process noises \citep{Dunik17,Zhang20}.
The trajectory as given by \eqref{eq:ft_Ft} may still be smooth even the target maneuvers \citep{Li19stf}. %; see the example studied in \citep{Li18FtC,}.
\SPrevisNew{A smooth trajectory basically indicates a small process noise in the classic Markov motion model. More formally, we have the following definition on the smoothness of a curve function}: {the T-FoT is twice continuously differentiable and is said to be $\beta$-smooth in a time-window $[k',k]$ if $\forall k'\leq t\leq k$,
\begin{equation}\label{eq:def_smoothness}
  \|\bigtriangledown^2 {F}(t;\mathbf{C})\|_\text{s}\leq \beta ,
\end{equation}
where $\|\cdot \|_\text{s}$ denotes the spectral norm (i.e. the largest singular value in different dimensions).}

{The constraint \eqref{eq:def_smoothness} upper bounds the acceleration of the target by $\beta$. %that the acceleration in any dimension of the positions is smaller than $\beta$.
In the case we consider, $\beta$ is relatively insignificant as compared with the uncertainty of measurements.}%As shown, the core difference to most existing tracking approaches is due to the use of the T-FoT $\mathbf{x}_t = {f}(t)$ in lieu of the Markov model $\mathbf{x}_k = f_k(\mathbf{x}_{k-1},\mathbf{w}_k)$., which is complete unknown. A key motivation and advantage of \eqref{eq:FoT_model} is that the kinematic model belong to the target is typically unknown to the tracker/sensor. In fact,

\subsection{Sliding Time-window LS T-FoT Fitting}
To account for a-priori model information, one may define a constraint or penalty/regularization factor $\Omega_{F}(\mathbf{C})$ within the optimization \citep{Li19stf,Li18MoC,Zhou21}.
%%For example, the end of the trajectory is known.
%such as
%${F}(t;\mathbf{C}) \in {\mathfrak{F}},$
%%\begin{align}
%% \begin{split} \label{eq:5}
%% {f}(t) \approx \underset{{F}(t;\mathbf{C})}{\text{argmin}} & \sum_{t=k'}^{k_2}\parallel \mathbf{y}_t-h_t({F}(t;\mathbf{C}),\bar{\mathbf{v}}_t)\parallel, \\
%% \text{s.t.} \hspace{0.5cm} & {F}(t;\mathbf{C}) \in {\mathfrak{F}},
%% \end{split}
%% \end{align}
%where $\mathfrak{F}$ represents a hypothesis space which the FoT is subject to, for example the space of spline functions of a finite order.
Constrained fitting has actually been intensively studied \citep{Boyd14}. Since we focus on non-cooperative targets in this work, we do neither assume constraints nor use any a-priori target models. %&composed of a set of component functions, e.g., polynomials% of a finite order
\SPrevis{Alternatively, a penalty factor $\Omega_{F}(\mathbf{C})$ may be used for measuring the disagreement between the fitting function and the constraint. For example, $\Omega_{F}(\mathbf{C}):=({F}(t;\mathbf{C})-\mathbf{x}_t)^{\text T}({F}(t;\mathbf{C})-\mathbf{x}_t)$ measures the mismatch between the fitting trajectory and the state $\mathbf{x}_t$ that the target passes by at time $t$. }
%Given guaranteed model constraint, the so-called optimal model error \citep{Mook88,Crassidis97} can be explicitly accounted for.

Obviously, the \textit{best} T-FoT ${F}(t;\mathbf{C})$ should minimize the approximate error $|\mathbf{\epsilon}(t)|$. But unfortunately, $\mathbf{x}_t$ is time-varying, unknown and is just what we want to estimate. We only have noisy measurements incoming in series at discrete-time-instants as shown in \eqref{eq:MeasurementModel}. % in the presence of false and missing data.
%The sensor is usually cooperative and so the measurement model \eqref{eq:MeasurementModel} is typically known or can be learned in advance. %\citep{Li17AGC,Li19stf}.
%Therefore, a
A natural idea is determining $\mathbf{C}$ as those which best fit the time series measurements %for which we consider the measurements
in a {sliding} time-window upto the current time $k$, denoted hereafter as
$$K:=[k', k]$$
where $k' = \text{max}(1,k-T)$, and $T$ is the length of the time-window.

Disregarding false and missing data issues temporally here, the T-FoT at time $k$ can be estimated by
\begin{equation} \label{eq:Ck_optimization}
\hat{\mathbf{C}}_{K} = \underset{\mathbf{C}}{\text{argmin}}  \sum_{i=k'}^{k}\mathcal{D}_i(\mathbf{C}) . %+ \lambda \Omega_{F}(\mathbf{C}).  % \| \mathbf{y}_t-\hat{\mathbf{y}}_t\|^2_{\Sigma^{-1}_{\mathbf{e}_t}}
\end{equation}
% where $\lambda>0$ controls the trade-off between the data fitting error $\mathcal{D}_i(\mathbf{C})$ and the model fitting error $\Omega_{F}(\mathbf{C})$.

In this work, the data fitting error $\mathcal{D}_i(\mathbf{C})$ is given in the LS sense using the Mahalanobis distance, i.e.,
\begin{equation} %\label{eq:D_Ck}
  \mathcal{D}_i(\mathbf{C}) := \| \mathbf{y}_i-\hat{\mathbf{y}}_i\|^2_{\Sigma^{-1}_{\mathbf{e}_i}} = (\mathbf{y}_i-\hat{\mathbf{y}}_i)^{\text T} \Sigma_{\mathbf{e}_i }^{-1}  (\mathbf{y}_i-\hat{\mathbf{y}}_i)  , \nonumber
\end{equation}
with a shorthand notation
\begin{equation}
\hat{\mathbf{y}}_i = h_i\left({F}(i;\mathbf{C})\right)+ \bar{\mathbf{v}}_i
\end{equation}
where $\bar{\mathbf{v}}_i$ gives the mean of the measurement noise $\mathbf{v}_i$ at discrete time $i$, \newrev{and the fitting error is given as
\begin{equation}\label{eq:fitting_err_e}
  \mathbf{e}_i := \mathbf{y}_i-\hat{\mathbf{y}}_i,
\end{equation}
which accounts for two sources of uncertainties including the measurement noise $\mathbf{v}_i$ and the T-FoT error $\mathbf{\epsilon}(i)$ and \newrev{$\Sigma_{\mathbf{e}_i}$ denotes the covariance of $\mathbf{e}_i$}; assumption or simplification will be made on the statistics of the fitting error; see \eqref{eq:approx_e}.}

\SPrevis{We have analyzed in \citep{Li19FoTClutter} that in a linear Gaussian system, the penalty factor can take into account the Markov dynamic model information if available such that the optimization amounts to that behind the KF. In other words, in the case of a deterministic Markov model (namely no process noise, $E(\mathbf{w}_k^{\text{T}}\mathbf{w}_k)=0$), the optimization function of the KF will reduce to that of the T-FoT-oriented approach.
%%To be more specific,
%The relevance of the optimization \eqref{eq:Ck_optimization} to that of the classic KF has been discussed in \citep[Section III.B]{Li19FoTClutter}. %
The advantage of the T-FoT formulation is that it does not assume a Markov model for modeling the movement of the target, nor imposes requirements on state temporal independence and on chronological sensor data. Different from existing modeling of the kinematic model error such as \citep{Gao21}, our formulation \eqref{eq:Ck_optimization} treats the model error in a flexible regularization-based manner.} %: how much $\Omega_{F}(\mathbf{C})$ affects the result depends on how much the a-priori model information is available.
%%In contrast, in the MME estimator \citep{Mook88,Crassidis97},
%Here, however, we consider as little as possible a-priori scenario information and so we assume no explicit constraint, %there is no constraint on the fitting function (namely $\Omega_{F}(\mathbf{C})$ vanishes from \eqref{eq:Ck_optimization}), the essence of
%the formulation \eqref{eq:Ck_optimization} reduces to estimating $\mathbf{C}_{K}$ as follows
% \begin{equation} \label{eq:Ck_optimization}
% \hat{\mathbf{C}}_{K} = \underset{\mathbf{C}}{\text{argmin}} \sum_{i=k'}^{k} \mathcal{D}_i(\mathbf{C}). % + \lambda \Omega_{F}(\mathbf{C}).
% \end{equation}
%which is unique if $\mathcal{D}_i(\mathbf{C})$ is convex.

Based on the T-FoT parameter estimate $\hat{\mathbf{C}}_{K}$ \SPrevis{conditioned on the measurements $\mathbf{y}_{k':k}$ received during time $[k', k]$}, the state at any time $t$ can be inferred as follows
\begin{equation} \label{eq:FoT_evaluation}
\hat{\mathbf{x}}_t|_{\mathbf{y}_{k':k}}={F}(t;\hat{\mathbf{C}}_{K}) \hspace{0.5mm}.
\end{equation}
More specifically, the inference is referred to as smoothing when $t <k$, as prediction or forecasting when $ t> k$ and as online filtering when $t=k$. We focus on filtering in this work, i.e., estimate of the target position will be updated whenever new measurements are available. %; we consider $t=k$ only in this work. %In this work, we concentrate on online tracking, i.e., the state to infer/filter is exactly for the time when the latest sensor data arrive.

\subsection{Online Estimating $\mathbf{C}$}
%\subsubsection{Linear Parameter Dependence}
For easy computation, linear parameter dependence is assumed in each dimension and smooth piecewise polynomial fitting function is then used,
i.e.,
\begin{equation} \label{eq:Ckpolynamial}
{F}^{(d)}(t; \mathbf{C})=c^{(d)}_0+ c^{(d)}_1t+\cdots+c^{(d)}_\gamma t^\gamma ,
\end{equation}
where $\gamma$ is referred to as the order of the fitting function and $d$ indicates the dimension in the position space.

%\begin{equation}
%\hat{f}(t) =
%\underset{{F}(t;\mathbf{C})}{\text{argmin}} \sum_{i=k'}^{k} \| \mathbf{y}_i-h_i({F}(t;\mathbf{C}),\bar{\mathbf{v}}_i)\|^2_{\Sigma^{-1}_{\mathbf{e}_i}} ,  % \| \mathbf{y}_i-\hat{\mathbf{y}}_i\|^2_{\Sigma^{-1}_{\mathbf{e}_i}}
%\end{equation}

{Parameters $c^{(d)}_0, c^{(d)}_1$ and $c^{(d)}_2$ indicate the initial position (at time $t=0$), velocity and acceleration of the target in dimension $d$, respectively.} As a rule of thumb, $\gamma = 1$ and $\gamma =2$ are suitable to model the (near) constant velocity (CV) and constant acceleration (CA) \citep{Li19stf}, respectively.
\newrev{In the case that the target has a definite originating position (\SPrevis{denoted by} $\mathrm{p}_0^{(d)}$ in dimension $d$), constant velocity ($\mathrm{v}^{(d)}$ in dimension $d$), or constant acceleration ($\mathrm{a}^{(d)}$ in dimension $d$), then the respective constraint $c^{(d)}_0= \mathrm{p}_0^{(d)}$, $c^{(d)}_1= \mathrm{v}^{(d)}$, or $c^{(d)}_2= \mathrm{a}^{(d)}$ needs to be satisfied in \eqref{eq:Ck_optimization}, respectively. Therefore, the parameters of the T-FoT approach are theoretically explicable, rendering the approach analytical. } % or \eqref{eq:Ck_optimization}.

%, given a properly defined length of the sliding time-window. That is, our approach poses nearly constant assumption on the velocity or acceleration in the piecewise time window. This is typically true and helpful in practical tracking; see the argument given on the traditional tracking framework \citep{Pilte17}.

%%\subsubsection{Recursive Updating}
\SPrevis{
For the above linear systems, % (i.e., the cost function is the linear function of $\mathbf{C}$%, which needs both measurement function $h_k$ and fitting function ${F}(t;\mathbf{C})$ being linear),
$\mathbf{C}$ can be calculated analytically and is unique; we will address this in the LS error sense in Section \ref{sec:LS_fitting}.
For a nonlinear system that corresponds to a non-convex optimization problem, one has to resort to iterative/numerical approximation methods %. % with guaranteed performance \citep{Chi19}. % \citep{Kelley99,Kanzow04}. %
%Iterative approximation methods work
on the basis of an initial guess of the parameters for iterative searching. %The parameters are obtained by progressive revision until the residuals to be minimized in \eqref{eq:Ck_optimization} do not decrease significantly in iterations or become lower than a threshold.
To speed up the optimization searching, it is important to start from the parameters $\hat{\mathbf{C}}_{[k'-1, k-1]}$ yielded at time $k-1$ for approaching the optimal $\hat{\mathbf{C}}_{K}$ at time $k$. This is reasonable since the two corresponding trajectories are actually overlapped in %yielded by the data in time-window $[k',k]$ and by those in time-window $[k-T+1,k+1]$ resonate due to
the piecewise $[k',k-1]$.
%, that is,
% \begin{equation} \label{eq:C_k}
% \mathbf{C}_k = \mathbf{C}_{k-1} + \mathbf{\rho}_{k-1},
% \end{equation}
%where $\mathbf{\rho}_{k-1}$ represents the minor difference between parameters of two adjacent time-windows. %, which does not need to be modelled or estimated.
%To speed up the optimization in the nonlinear measurement system, we set the parameters $\hat{\mathbf{C}}_{[k'-1, k-1]}$ yielded at time $k-1$ as the initial parameters to start from for searching for the optimal parameters $\hat{\mathbf{C}}_{K}$ at time $k$. This is reasonable because the trajectory functions yielded by the data in time-window $[k',k_2]$ and by those in time-window $[k'+1,k+1]$ are similar due to the common data in $[k'+1,k_2]$.
This amounts to assuming a recursion on the latent T-FoT parameters which is critical for ensuring online implementation for non-convex optimization.}
%However, it has been very recently pointed out by Pacholska \textit{et al.} in a nice contribution \citep{Pacholska20} that, the above iterative solver may only converge to a local minimum. They %further solve the problem as referred differently to \textit{trajectory recovery}, in closed-form and
{Recently, Pacholska \textit{et al.} \citep{Pacholska20} relaxed the problem and addressed the sufficient conditions for the guaranteed fitting optimality/uniqueness, where the measurement considered in particular is the distance between the sensor and the target.} \newrev{Even more recently, the polynomial fitting has been used for fitting measurement in tracking \citep{Tian22}.} \SPrevis{A sensor selection approach based on the T-FoT approach has been proposed \citep{Li22SensorSelection} for online activating a finite number of sensors in a sensor network with communication bandwidth constraints.}

{As addressed so far, %neither \eqref{eq:Ck_optimization} nor \eqref{eq:Ck_optimization} takes
existing T-FoT-based approaches have not take
into account false and missing data at unknown rates/ratios, or the birth and death of the target, which are essentially lying in the core of realistic target tracking.}

\section{Model and Scenario Assumptions}
\label{sec:model}
There is no more than one target in the scenario. If it exists and is detected, the position measurement %(e.g., received by using a camera)
%The position measurement
\SPrevis{$\mathbf{y}_{k} := \begin{bmatrix} y^{(1)}_{k} \\ y^{(2)}_{k} \end{bmatrix} $ is given by
\begin{align}\label{eq:Position_Measure}
   \mathbf{y}_{k} = \begin{bmatrix} p_{x,k} \\ p_{y,k} \end{bmatrix} + \mathbf{v}_{k} ,
\end{align}}
where $[p_{x,k}, p_{y,k}]^\text{T}$ is the position of the target, and $\mathbf{v}_{k}$ is the measurement noise. % of which the covariance is denoted as $\mathbf{R}_k$.

%The noisy position measurements are available with a fixed detection probability $p^\text{D}(\mathbf{x}_k) $.
We also consider the position-relevant measurements that can be converted to positions at each scan, i.e., the measurement function is injective \citep{Li16O2} or multiple sensors being used \citep{Li17clustering,Li18FtC}.
For example, the range-bearing measurement model %for an active radar located at $[s_x,s_y]^{\text{T}}$ %considered in our second simulation
can be written as follows
\begin{equation} \label{eq:range-bearing-measurement}
\begin{bmatrix} r_{k} \\[.5mm] \theta_{k}  \end{bmatrix}   = \begin{bmatrix}
\sqrt{ p_{x,k}^2 + p_{y,k}^2} \,\,\\
\vspace{0.5mm}
\tan^{-1}\!\Big( \frac{ p_{y,k} }{ p_{x,k}} \Big)
\end{bmatrix}
+ \mathbf{v}_{k} ,  %\begin{bmatrix} v_{{r},k}  \\[0.5mm] v_{\theta,k} \end{bmatrix},
\end{equation}
%which can be converted to the position measurements by
%\begin{equation} \label{eq:rang-bearing-conversion}
%\mathbf{\tilde{y}}_{k}:=\begin{bmatrix} \tilde{y}^{(1)}_{k} \\ \tilde{y}^{(2)}_{k} \end{bmatrix} = \begin{bmatrix} \lambda_\theta r_{k} \mathrm{cos}(\theta_{k}) \\[.5mm] \lambda_\theta r_{k} \mathrm{sin}(\theta_{k}) \end{bmatrix},
%\end{equation}
%where $ \lambda_\theta =  \text{exp}(-\frac{\Sigma_\theta}{2}) $.
which can be converted to the position measurements by
\SPrevis{\begin{equation} \label{eq:rang-bearing-conversion}
\mathbf{y}_{k} = \begin{bmatrix} r_{k} \mathrm{cos}(\theta_{k}) \\[.5mm] r_{k} \mathrm{sin}(\theta_{k}) \end{bmatrix} .
\end{equation}
}
The above conversion has a bias in the mean of the converted measurement, % \citep{Bar-Shalom01}
which can be removed by multiplicating a factor either $\text{exp}(-\frac{\Sigma_\theta}{2})$ or $\text{exp}(\frac{\Sigma_\theta}{2})$ \citep{Bordonaro14}. %\revis{
\SPrevis{Further on, the converted covariance $\Sigma_{\mathbf{{y}}_{k}}$ can also be calculated numerically via Monte Carlo sampling \citep{Li16O2} for any type of noises. Discussion on the information gain/loss due to measurement conversion can be found in \citep{Lan20conversion} and the reference therein.} % or analytically \citep{Bordonaro14} for Gaussian noises. % based on $\Sigma_{\mathbf{{y}}_{k}}$ (or to say, $\Sigma_{\mathbf{v}_{k}}$) and $\mathbf{{y}}_{k}$. %In our simulation, we use the so-called Modified Unbiased Converted Measurement (MUCM) method \citep{Bordonaro14}.
 % can be equally convert to the above position measurement model, as is done in our simulation. %Further discussion on this can be found in \citep{Li16O2, Li17clustering, Li17MSDC}.
% We use this  in our second simulation in Section~\ref{sec:nonli-Sim}.%This serves as an overfit case of T-FoT fitting where the fitting function (although not explicitly given) works specifically on the sensor data.

The measurements at time $k$ are denoted by a %random finite
set $\mathbf{Y}_k$ \SPrevis{which are composed of both real measurement of the target (if detected) and clutter at time $k$}. The following assumptions are required in our work.
\begin{itemize}
  \item[A1.] The measurement noise $\mathbf{v}_{k}$ which might not be Gaussian has a zero-mean and variance $\Sigma_{\mathbf{v}_{k}}$ given a-priori.
  \item[A2.] Clutter is independently generated at different times, (near-)uniformly distributed over the surveillance region and independent to the measurement of the real target;
  %\item sufficient observability is guaranteed as measurements are converted to position measurement in the state space.
  \item[A3.] Clutter rate $r_c$ is relatively low as required by an algebraic manipulation analysis given in {Appendix \ref{appendix_Clutter}}. In our simulations, $r_c<5$.
  \item[A4.] The target detection probability $p^\text{D}$ at any time is fairly high, e.g., larger than threshold $p^\text{D} =0.9$.
\end{itemize}

{Further on, as the particular scenario in which the T-FoT approach performs best, the trajectory of the target is ``smooth" having a minor $\beta$ in \eqref{eq:def_smoothness}, i.e., the target moves with relevantly insignificant acceleration and process noise as compared with its velocity.} This together with the above assumptions are representative and match the real case of many realistic problems particularly for space target tracking (e.g., air traffic/aviation supervision). %This is indeed an
%\begin{itemize}
%  \item[A4.] %(while maneuvers may occur sometimes)
%  %and the impact of the process noise or unknown input to the position of the taget is much smaller as compared with the average movement of the target in one sensing period. %All these ensure. To ensure this,.
%  %\item[A5.] In the view of the kinematic model for describing the dynamics of the target as in \eqref{eq:Markov}, the process noise is insignificant, i.e., $\Sigma_{\mathbf{w}_k}$ is minor or even zero.
%\end{itemize}

%As shown the above assumptions use two probability thresholds $p_r, p_D$, which can be directly set as $p_r=0.9, p_D=0.6$.

\section{Proposal: T-FoT Initiation, Maintenance and Termination} \label{sec:proposal}

This section addresses the details of our approach for T-FoT initiation, maintenance and termination, respectively.

\subsection{Trajectory Initiation} \label{sec:LS_fitting} %\revis{--- This section is new}
The lack of a-priori information about the target birth tangled with false and missing data lead to a significant challenge for initializing the tracker.
%The lack of information about target birth make it challenging for target detection. To solve this, we
In this work, we propose to cluster the time-series measurements of the sliding time-window $K$ for detecting the birth of the target. This requires that the real measurements of the target in successive scans are significantly closer to each other on average than those between different clutter points and between the target and clutters. This is guaranteed by the above assumptions A1$-$A3.
%It is simply impossible to detect the target immediately from a set of cluttered data (which may not include the real data) at a single sensing scan unless the target generates multiple measurements. The only solution is to jointly consider the measurements over successive scans (a time-window).
%we can cluster .  %the target trajectory once detecting its appearance.
To this end, the density-based clustering method called DBSCAN \citep{Ester96} is readily available that distinguishes regions of high data density from low-density regions and does not need to be specified with the number of clusters in advance. For this purpose, %where $k'=\mathrm{max}(1,k-T)$ and $k$ is the newest time,
two key parameters are required as to be addressed next: 1) the neighborhood radius $\varepsilon$: the maximum distance between two neighbor measurements in the cluster, %which is the minimum Mahalanobis distance between any data-point and its nearest neighbor in the cluster,
and 2) the minimum cluster size $T_\text{s}$. %: the target will only be declared if the cluster has a size larger than $T_\text{s} \leq T$. %Once a target is detected via the clustering operation, we apply a (weighted) LS fitting to the clustered measurements.

In addition, the following \textit{point-target constraint} needs to be followed to guide the clustering: data-points in the cluster corresponds to different time-instants. This constraint can be formulated as a cannot link rule \citep{Li17MSDC}: the measurements generated at the same time-instant cannot not belong to the same cluster.
However, we note that, it is possible yet rare that two clusters are formed at the same time, at least one of which must be false alarm. In this case, we increase both the minimum cluster size $T_\text{s}$ and the length of the time-window gradually and re-do the clustering till there is only one cluster to meet the single-target assumption.

\subsubsection{Neighborhood radius $\varepsilon$}
The distribution of real measurements of the target depends on two factors: 1) the measurement noise; \newrev{see the analysis given in Appendix \ref{appendix_Dist_Gauss_Variable}}, and 2) the velocity of the target \citep{Musicki13,Mallick15} to account for the movement of the target in one sensing period. In the extreme case that both the initial speed of the target and the statistics of the measurement noise are unknown, one have to learn $\varepsilon$ from the data.
%\footnote{We have earlier proposed an unsupervised clustering algorithm for the similar purpose in \citep{Li17MSDC} which has been further improved in \citep{Fan19c4f}; see also some other works such as \citep{Rodriguez14science} for online learning a similar parameter called cut-off distance. Here, it is necessary to note in these applications, it is implicitly assumed that there is at least one cluster/target in the data set. This does not hold in our approach as the target may not present. }. %In our present study, however, we assume that the covariance of the zero-mean measurement noise is given a-priori and we do not consider a high initial target speed.
%Based on assumptions A1~A4, it is reasonable to believe that the distance between a clutter point and the target or between any two clutter points are significantly larger than that between the measurements of the target at successive time-instants.
The problem of whether two measurements are generated by the same target can be modelled
as a composite binary-hypothesis testing problem as follows
\begin{equation}\label{Bernoulli-RFS}
\left\{ {\begin{array}{l} \nonumber
	\mathcal{H}_0: \text{at least one measurement is clutter} \\
	\mathcal{H}_1: \text{both measurements are from the target}
	\end{array}} \right.
\end{equation}

%We evaluate the distance between measurements by the Mahalanobis distance (which has a clear meaning as addressed in Appendix B) and specify
The hypothesis testing is carried out by comparing the Mahalanobis distances between measurements with the neighborhood radius $\varepsilon$, i.e.,
\begin{equation} \label{eq:tau_1}
 \| \mathbf{y}_i-\mathbf{y}_j\|^2_{\Sigma^{-1}_{\mathbf{v}_i}}   \underset{{\mathcal{H}_1}}{\overset{{\mathcal{H}_0}}{\gtreqless}}   \varepsilon := \tau^2_1 \hspace{0.5mm}. %\mathcal{D}({\mathbf{y}}_1,{\mathbf{y}}_2)
\end{equation}
%where $\mathcal{H}_0, \mathcal{H}_1$ are the null and alternative hypotheses that measurements $\mathbf{y}_i, \mathbf{y}_j$ are generated from the target, respectively.

%Straightforwardly, the distribution of the
Mahalanobis distance between two independently identically distributed (IID) variables is analyzed in Appendix \ref{appendix_Dist_Gauss_Variable}. However, the target measurements at two time-instants are not IID due to the varying of the target state. To compensate for the velocity of the target, we use a threshold $\tau_1$ in \eqref{eq:tau_1} larger than necessarily required when the target is stationary as analyzed in Appendix B, e.g., $\tau_1=3$ in our simulation. Obviously, this works only when the initial velocity of the target is small. %where ${\mathbf{y}}_1,{\mathbf{y}}_2$ represent the concerning two measurements. % is also known as the Mahalanobis distance between the measurement ${\mathbf{y}}_k$ and its neighbor ${\mathbf{y}}^{\text{N}}_k$ as shown in \eqref{eq:D_Ck}.
 %both the clutter rate and the initial speed of the target are not so high so that

%``roughly accurate''\footnote{By this, we mean that the information does not need to be exactly accurate. Let us say, the estimate error relevant to the truth is less than 20\%.} estimates of the initial and the measurement covariance, as well as, can lead to a final rough estimate of $\varepsilon$. %Indeed, we do so in our simulation.

\subsubsection{Minimum cluster size $T_\text{s}$} The proposed track initiation follows a ``$m$ out of $n$" logic \citep{Castella76,Bar-Shalom89,Hu97,Worsham10} but our approach is free of the knowledge of the process noise statistics, i.e., a cluster/track is confirmed if and only if at least $T_\text{s}$ detections are clustered over $K$ scans.
We now address the choice of $T_\text{s}$. First of all, it must be larger than the number of parameters of the fitting function to avoid underfit and is no larger than the average number of detections in the time-window, that is,
\begin{equation} \label{eq:scope_Ts}
 T_\text{s} \in [\gamma+1, Tp^\text{D}].
\end{equation}
%A typical choice is $T_\text{s}=3$. % or $T_\text{s}=4$. %we use $T_\text{s}=4$ when $\gamma=2$ (for a potentially CA model) and $T_\text{s}=3$ when $\gamma=1$ (for a potentially CV model).

%Then, we initialize the trajectory by fitting these clustered measurements otherwise increase the length of the time window with the increase of time $k$ and redo clustering till a target is detected.

Note that as long as the \textit{point-target constraint} holds, target detection can be earliest made $t \geq T_\text{s}$ filtering steps after the target appears. So, $T_\text{s}$ indicates how much latency/delay we have to endure at least for identifying the target; analytical results on the statistics of the detection delay have been given in \citep{Castella76,Bar-Shalom89,Hu97,Worsham10}. %In this sense, smaller $T_s$ is better. %In general, a smaller $T_\text{s}$ implies a shorter detection delay but also a higher risk of causing false alarms.
Assume that the probability for all clutter points generated at a particular sampling scan falling further than the threshold (corresponding to the radius of the cluster) to any given point in the surveillance area is $p_r$, c.f. \eqref{eq:circleProb_rc}; \SPrevis{see the analysis given in Appendix A}. Then, the probability for causing a false alarm (FA) due to the clustering of at least $T_s$ clutter points (obtained at different time-instants) %around a point in the surveillance area
can be approximately estimated by %, if $T_s \leq T-1$,
\begin{align}
p_\text{FA}(T_s) & \approx \sum_{t=T_s}^TC_T^{t}(1-p_r)^t p_r^{T-t} \nonumber \\
& = C_T^{T_s}(1-p_r)^{T_s} p_r^{T-T_s}  + p_\text{FA}(T_s+1) , \label{eq:pFA}
\end{align}
where $C_m^n$ stands for the number of combinations of $n$ elements taken from a set of size $m$.

As shown, $p_\text{FA}(T_s)$ decreases with the increase of $T_s$, that is, larger $T_s$ implies smaller FA rate but also greater target-detection delay. A tradeoff is required here for which $3, 4$ are our recommendation.

\subsubsection{T-FoT initiation Based on Weighted LS fitting}
Denote the measurements in the confirmed cluster as $\{\mathbf{y}_i\}_{i \in \tilde{K}} \subseteq \{\mathbf{Y}_i\}_{i=k'}^k$ where $\tilde{K} \subseteq \{k', k'+1, \dots, k\}$. %$\tilde{K}$ may have identical elements\footnote{This is simply due to multiple measurements generated close to the potential track of the target and we cannot identify which is real measurement. However, the fitting can be the same applied for this case. }, of which all the elements belong to $[k', k]$.
%(Measurements are converted to positions if they are not originally position measurements. With a bit notational abuse, they are still denoted by $\{\mathbf{y}_i\}_{i \in \tilde{K}}$ for notation simplicity.)
According to \eqref{eq:Ckpolynamial}, we assume the T-FoT in each position dimension as follows
\begin{equation} \label{}
\begin{bmatrix} y^{(1)}_{i} \\ y^{(2)}_{i} \end{bmatrix} = \begin{bmatrix} c^{(1)}_0, c^{(1)}_1, \cdots,c^{(1)}_\gamma \\ c^{(2)}_0, c^{(2)}_1, \cdots,c^{(2)}_\gamma \end{bmatrix}  \begin{bmatrix} 1 \\ i \\ \cdots \\ i^\gamma \end{bmatrix}  + \begin{bmatrix} {e}^{(1)}_i \\ {e}^{(2)}_i  \end{bmatrix},
\end{equation}
where ${e}^{(d)}_i$ is the fitting error at time $i \in \tilde{K}$ in dimension $d=1,2$; see \eqref{eq:fitting_err_e} and \eqref{eq:approx_e}. %, which accounts for two types of uncertainty including the measurement noise $\mathbf{v}_k$ and the T-FoT model error $\mathbf{\epsilon}(t)$.

The T-FoT parameters %$\mathbf{C}_k := \{C^{(1)}_k,C^{(2)}_k\}$ with components
$\mathbf{C}^{(d)}_{\tilde{K}} :=\left[c^{(d)}_0, c^{(d)}_1, \cdots,c^{(d)}_\gamma\right]^\text{T}, d=1,2$ %C^{(2)}_k =\{ c^{(2)}_0, c^{(2)}_1, \cdots,c^{(2)}_\gamma \}$
can be estimated in the LS sense, i.e.,
\begin{equation} \label{eq:hat_bf{C}_k^d}
 \hat{\mathbf{C}}^{(d)}_{\tilde{K}}  = \underset{ c^{(d)}_0, c^{(d)}_1, \cdots,c^{(d)}_\gamma }{\text{argmin}} \sum_{i \in \tilde{K}} \|{y}^{(d)}_i- \hat{y}^{(d)}_i \|^2_{\Sigma^{-1}_{{e}^{(d)}_i}} .
\end{equation}
%which factorizes \eqref{eq:Ck_optimization}. % under no model constraint.

%Fortunately, \eqref{eq:hat_bf{C}_k^d} can be solved analytically.
Given that $\mathbf{A}^T\Sigma_{\tilde{K}}^{-1}\mathbf{A}$ is non-singular (which is easy to be satisfied when $|\tilde{K}|>\gamma$ and when $\tilde{K}$ has no identical elements), the exact solution to \eqref{eq:hat_bf{C}_k^d} is given by
\begin{equation}\label{eq:Exact_solution_for_Ck}
 \hat{\mathbf{C}}^{(d)}_{\tilde{K}}   = (\mathbf{A}^\text{T}\Sigma_{\tilde{K}}^{-1}\mathbf{A})^{-1} \mathbf{A}^\text{T}\Sigma_{\tilde{K}}^{-1}\mathbf{Y}_{\tilde{K}}^{(d)} ,
\end{equation}
where $\Sigma_{\tilde{K}} = %\mathrm{diag} (\Sigma_{{e}^{(d)}_{k'}}, \Sigma_{{e}^{(d)}_{k'+1}}, \cdots, \Sigma_{{e}^{(d)}_{k}})$$
\text{Cov}\big(\mathbf{e}_{\tilde{K}}^{(d)} %(\mathbf{e}_{\tilde{K}}^{(d)})^\text{T}
\big)$, $\mathbf{e}_{\tilde{K}}^{(d)} = [{{e}^{(d)}_{k'}}, {{e}^{(d)}_{k'+1}}, \cdots, {{e}^{(d)}_{k}}]^{\text{T}}$ %$\Sigma = \text{diag}{(\Sigma_{\mathbf{y}^{(d)}_{k'}}, \cdots,\Sigma_{\mathbf{y}^{(d)}_{k}})}$ due to the independence between measurements at different times in $\tilde{K}$,
and
\[
\mathbf{A} =
\left[ \begin{array}{cccc}
1 & k' & \cdots & (k')^\gamma \\
1 & k'+1 & \cdots & (k'+1)^\gamma \\
 & \vdots \\
1 & k & \cdots & (k)^\gamma \\
\end{array} \right]
%\begin{bmatrix} 1, k', \cdots,(k')^\gamma \\ 1, k'+1, \cdots,(k'+1)^\gamma \\ \cdots \\ 1, k, \cdots,k^\gamma \end{bmatrix}
, \mathbf{Y}_{\tilde{K}}^{(d)} = \begin{bmatrix} {{y}^{(d)}_{k'}}\\ {{y}^{(d)}_{k'+1}}\\  \vdots \\ {{y}^{(d)}_{k}} \end{bmatrix}. %, \mathbf{e}^{(d)} = \begin{bmatrix} {{e}^{(d)}_{k'}}\\ \cdots \\ {{e}^{(d)}_{k}} \end{bmatrix},
\]

%\begin{corollary}
%Assume that fitting errors $\mathbf{e}^{(d)}$ have zero mean, are homoscedastic and uncorrelated between distinct terms, namely $\mathbf{E}(e^{(d)}_i) = 0, \Sigma_{e^{(d)}_i} = \sigma^2,  i=1,\cdots,n$ and $\text{Cov}(e^{(d)}_i, e^{(d)}_j) =0, \forall i \neq j$. Then, the Gauss-Markov theorem \citep[pp.34]{Kariya04} indicates that the above LS estimate $\hat{\mathbf{C}}^{(d)}_{\tilde{K}} $ as given in \eqref{eq:hat_bf{C}_k^d} is the best linear unbiased estimate of $\mathbf{C}_k$. %the covariance of the estimate is given
%\end{corollary}
%\begin{corollary}
Assume fitting errors %$\mathbf{e}_{\tilde{K}}^{(d)}$
having zero mean, namely $\mathbf{E}(e^{(d)}_i) = 0, \forall i \in \tilde{K}$. Then, the Gauss-Markov theorem \citep[pp.34]{Kariya04} \citep[pp.86]{Luenberger68} indicates that the LS estimate $\hat{\mathbf{C}}^{(d)}_{\tilde{K}} $ as given in \eqref{eq:Exact_solution_for_Ck} is the best linear unbiased estimate (BLUE) of $\mathbf{C}^{(d)}_{\tilde{K}}$, i.e., $\mathbf{E}(\hat{\mathbf{C}}^{(d)}_{\tilde{K}}) = \mathbf{C}^{(d)}_{\tilde{K}}$, and
\SPrevis{
\begin{equation}\label{eq:covC}
  \mathbf{Cov}\big(\hat{\mathbf{C}}^{(d)}_{\tilde{K}}\big) = \big(\mathbf{A}^\text{T}\Sigma_{\tilde{K}}^{-1}\mathbf{A}\big)^{-1}.
\end{equation}}
%the covariance of the estimate is given
%\end{corollary}
%\begin{proof}
See the proof for Theorem 2.1 of \citep[pp.34]{Kariya04} and for Theorem 1 of \citep[pp.86]{Luenberger68}. The BLUE is also referred to as minimum variance unbiased estimator in \citep{Luenberger68}.
%\end{proof}

%If the fitting error $\mathbf{e}^{(d)}$ is multinormal, with zero mean and covariance $\sigma^2\mathbf{I}_n$, it follows from the Lehmann-Scheffe theorem that $\hat{\mathbf{C}}^{(d)}_{\tilde{K}}$ is the (almost surely unique) uniformly minimum variance unbiased estimator of ${C}^{(d)}_k$ \citep[Ch.2]{Lehmann98}.

\begin{figure}
\centering
\includegraphics[width=10cm]{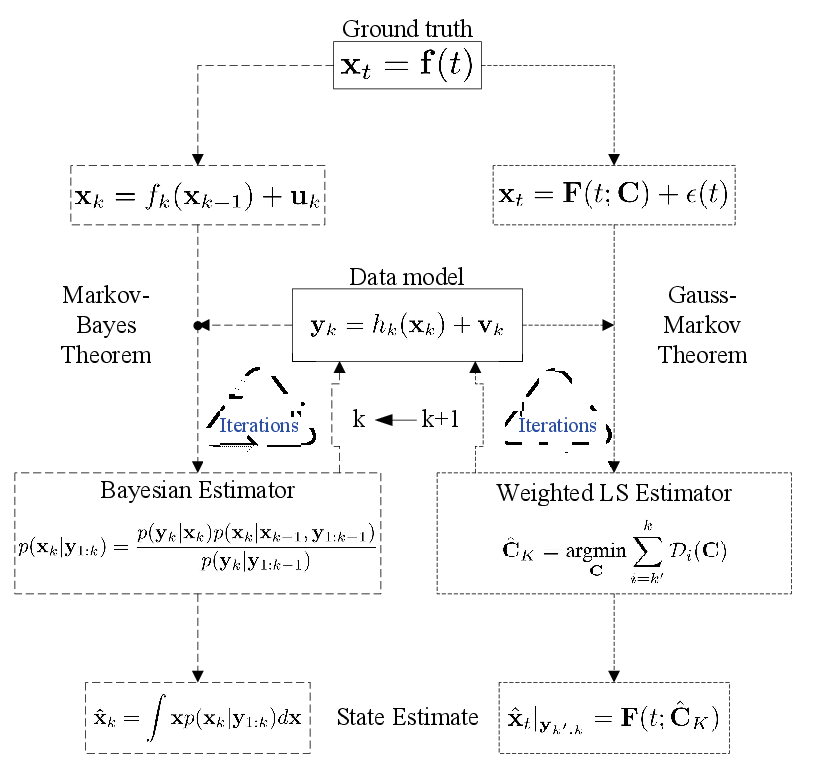}
\caption{\revis{Comparison between the Markov-Bayes estimator based on the SSM and the weighted LS estimator based on the T-FoT.}}\label{fig:ssmFoT}
\vspace{-1mm}
\end{figure}

\revis{The key difference between our proposed T-FoT estimator and the classic Markov-Bayes estimator can be illustrated in Fig~\ref{fig:ssmFoT}. Both approaches basically fit the real target trajectory with a prescribed model such that the output of the model best fits a series of measurements. To this end,}
%Overall,
the T-FoT-based LS fitting approach fits the ground truth by a continuous-time curve function while the KF assumes a Markov-jump model. \revis{They are based on the Markov-
Bayes theorem and the Gauss-Markov theorem, respectively.} %This leads to the difference between the respective BLUEs they admit, governed by different models and theorems.
Arguably speaking, they suit non-cooperative and cooperative target, respectively. More discussion on the relationship of the LS approach and the KF-type estimator are available in \citep{Sorenson70,Sayed94,Humpherys10,Li19FoTClutter}. %That is, ased on different models, both BLUE approaches hinge on different theorems and lead to different results.

\subsubsection{Approximate Fitting Error $\mathbf{e}_{i} $}
Obviously, $\mathbf{e}_{i} $ plays a key role in either \eqref{eq:Ck_optimization} or \eqref{eq:Exact_solution_for_Ck}. As addressed, it accounts for two sources of uncertainties including measurement noise $\mathbf{v}_i$ and T-FoT error $\mathbf{\epsilon}(i)$ which, however, is generally unknown. Since the cost function uniformly accounts for the fitting errors at different time instants in the concerning time-window $K$, a reasonable assumption is that the fitting errors $\mathbf{\epsilon}(i)$ is insensitive to time $i$. Then, %a reasonable surrogate of the fitting error is given by
 \begin{equation}\label{eq:approx_e}
    \Sigma^{(d)}_{\mathbf{e}_i} %\approx \mathrm{const} \times
    \propto
\Sigma^{(d)}_{\mathbf{v}_i}.
 \end{equation}
This substitution brings much convergence for calculation %as the constant can be simply removed from the optimization
in either \eqref{eq:Ck_optimization} or \eqref{eq:Exact_solution_for_Ck}. If further $\Sigma_{\mathbf{v}_i}$ is time-invariant, the weighted LS reduces to the ordinary LS.

\subsection{Trajectory Maintenance} %\revis{--- I tried the approach of ``predictive LS'' which, however, does not perform better result than the current approach (and also it seems that the predictive LS does not accommodate missed detections). }
The main idea of our trajectory maintenance algorithm is to update \eqref{eq:Exact_solution_for_Ck} with time $k$ increases. The key challenge is still from the misdetection and clutter. To enable the sliding time-window fitting, one has to identify the real measurements of the target (and their corresponding time-instants) from the clutter. %, by two major steps over a sliding time window, for which we present a predictive LS method; Second, update the T-FoT upon identifying any new real measurements.
To be more specific, the proposed trajectory maintenance scheme \SPrevis{comprises the following three steps}.

%\subsubsection{One-step prediction}
%Following \eqref{eq:FoT_evaluation}, the estimate of the state of the target at time $k+1$ can be inferred by using the T-FoT parameter $\hat{\mathbf{C}}_{[k', k]}$ obtained at time $k$ as follows
%\begin{equation} \label{eq:One-step_Prediction}
%\hat{\mathbf{x}}_{k+1|k}={F}(k+1;\hat{\mathbf{C}}_{[k', k]}) .
%\end{equation}
%%which can be viewed as a one-step state-prediction.
%
%\subsubsection{Calculation of the pseudo measurement}
%The pseudo measurement is calculated by using the measurement function and the predicted state \eqref{eq:One-step_Prediction} by
%\begin{equation} \label{eq:Pseudo_Measurement}
%\hat{\mathbf{y}}_{k+1} = h_{k+1}\left(\hat{\mathbf{x}}_{k+1|k}\right) + \bar{\mathbf{v}}_{k+1} .
%\end{equation}
%%where $\bar{\mathbf{v}}_{k+1}$ is the mean of the measurement noise at time $k+1$.

\subsubsection{Measurement one-step prediction}
Given the identified real measurements $\{\mathbf{y}_i\}_{i \in \tilde{K}} \subseteq \{\mathbf{Y}_i\}_{i=k'}^k$ where $\tilde{K} = \{\tilde{k}', \cdots, \tilde{k}\} \subseteq \{k', k'+1, \dots, k\}$, the measurement of the target at time $k+1$ can be predicted following the approach of ``predictive LS'' \citep{Rissanen86pls,Niedzwiecki17}, if there are no missed detections. % as follows
%\begin{equation} \label{eq:Pseudo_Measurement_PLS}
%\hat{\mathbf{y}}_{k+1} = \sum_{i \in \tilde{K}} \mathbf{a}_i \mathbf{y}_i.   %+ \text{error}
%\end{equation}
%Denote by $\mathbf{A}_k^{(d)} = [\mathbf{a}_{\tilde{k}'}^{(d)} ,\cdots, \mathbf{a}_{\tilde{k}}^{(d)} ]^T$ and by $\mathbf{Y}_k^{(d)}  = [\mathbf{y}_{\tilde{k}'}^{(d)} ,\cdots, \mathbf{y}_{\tilde{k}}^{(d)}]$. Then, \eqref{eq:Pseudo_Measurement} can be reformulated as follows
%\begin{equation} \label{eq:PM_d}
%\hat{\mathbf{y}}_{k+1}^{(d)}  =  \mathbf{Y}_k^{(d)} \mathbf{A}_k^{(d)}. % + \text{error}^{(d)}
%\end{equation}
%Then, the well-known exponentially weighted LS
%(EWLS) method can be used for estimating $\mathbf{A}^{(d)}$ from $\{\mathbf{y}_i\}_{i \in \tilde{K}}$. To achieve the effect of forgetting or discounting history data, a forgetting factor $\lambda$, e.g., 0.98, can be applied, resulting in the following EWLS estimator
%\begin{equation} \label{eq:PM_d}
%\hat{\mathbf{A}}^{(d)} =  \underset{\mathbf{A}^{(d)}}{\text{argmin}} \sum_{i =1}^{|\tilde{K}|-1} \lambda^{k-\tilde{K}(i)}\|{\mathbf{y}}_{\tilde{K}(i+1)}^{(d)} - \hat{\mathbf{y}}_{\tilde{K}(i)}^{(d)}  \|^2.
%\end{equation}
%Prediction error of the predictive LS has been analyzed in \citep{Niedzwiecki17} and is omitted here.
Alternatively and to accommodate potentially missed detection at any time instant, following \eqref{eq:FoT_evaluation}, the estimate of the state of the target at time $k+1$ can be inferred by using the T-FoT parameter $\hat{\mathbf{C}}_{\tilde{K}}$ obtained at time $k$ as follows
\begin{equation} \label{eq:One-step_Prediction}
\hat{\mathbf{x}}_{k+1|k}={F}(k+1;\hat{\mathbf{C}}_{\tilde{K}}) .
\end{equation}
%which can be viewed as a one-step state-prediction.
Then, a pseudo measurement is calculated by using the measurement function and the predicted state \eqref{eq:One-step_Prediction} by
\begin{equation} \label{eq:Pseudo_Measurement}
\hat{\mathbf{y}}_{k+1} = h_{k+1}\left(\hat{\mathbf{x}}_{k+1|k}\right) + \bar{\mathbf{v}}_{k+1} .
\end{equation}
%where $\bar{\mathbf{v}}_{k+1}$ is the mean of the measurement noise at time $k+1$.

%where $\bar{\mathbf{v}}_k$ is the mean of the noise.

\subsubsection{Distinguishing real measurement from clutter}
Denote the measurement set with $I$ elements received at time $k+1$ as $\mathbf{Y}_{k+1} = \{\mathbf{y}^1_{k+1},\mathbf{y}^2_{k+1},...,\mathbf{y}^I_{k+1}\}$.
A potentially real measurement will be identified based on their (Mahalanobis) distance to the pseudo measurement as follows:
\begin{equation} \label{eq:ML_Measurement}
\tilde{\mathbf{y}}_{k+1} = \underset{\mathbf{y}^i_{k+1} \in \mathbf{Y}_{k+1}}{\text{argmin}} \| \mathbf{y}^i_{k+1} -\hat{\mathbf{y}}_{k+1} \|_{\Sigma^{-1}_{\mathbf{e}_k}} % p(\mathbf{y}^i_k|\hat{\mathbf{x}}_{k|k-1}) .
\end{equation}
which yields the same result %as the minimization of the Mahalanobis distance but computes faster and
\newrev{as the maximal likelihood criterion when the fitting error belongs to the exponential family \citep{Streit94}.}
%Here, assume that the measurement noise is additive zero-mean and Gaussian distributed, and that the prediction $\hat{\mathbf{y}}_k$ is unbiased, the pseudo likelihood function is defined as
%\begin{equation}\label{eq:ML_likelihood}
%  p(\mathbf{y}^i_k|\hat{\mathbf{x}}_{k|k-1}) := \mathcal{N}(\mathbf{y}^i_k; \hat{\mathbf{y}}_k, \mathbf{P}),
%\end{equation}
%where $\mathbf{P}$ is chosen as $\mathbf{P} = \Sigma_{\mathbf{y}_k}$

To identify whether $\tilde{\mathbf{y}}_{k+1}$ is a real measurement of the target at time ${k+1}$ or actually there is no real measurement at all (namely the target is missed in detection), %a reasonable criterion is the %\textit{normalized estimation error squared (NEES)}, also known as the
%we compare the pseudo likelihood in \eqref{eq:ML_likelihood} with a threshold $\mathcal{N}(\tau_2\mathbf{I}_{D_\mathbf{y}};\mathbf{0},\mathbf{I}_{D_\mathbf{y}})$.
the Mahalanobis distance $\|\tilde{\mathbf{y}}_{k+1} -\hat{\mathbf{y}}_{k+1} \|^2_{\Sigma^{-1}_{\mathbf{v}_{k+1}}} $ can be used \citep{Gao22}.
That is, if it is not larger than threshold $\tau_2^2$, %$\mathcal{N}(\tau_2\mathbf{I}_{D_\mathbf{y}};\mathbf{0},\mathbf{I}_{D_\mathbf{y}})$,
$\tilde{\mathbf{y}}_{k+1}$ will be identified as the real measurement otherwise, misdetection is identified. That is
\begin{align} \label{eq:gating_y_NN}
\begin{cases}
\mathbf{y}_{k+1} := \tilde{\mathbf{y}}_{k+1}, & \text{if} \hspace{1mm} \| \tilde{\mathbf{y}}_{k+1} -\hat{\mathbf{y}}_{k+1} \|^2_{\Sigma^{-1}_{\mathbf{v}_{k+1}}} \leq \tau_2^2 \\
\{\mathbf{y}_{k+1}\} := \emptyset, & \text{otherwise}.
\end{cases}
\end{align}
\SPrevis{
However, when the target trajectory is not so smooth (corresponding to a large $\beta$ in \eqref{eq:def_smoothness}), the one-step prediction based on \eqref{eq:One-step_Prediction} may be inaccurate. For example, when the target makes a sudden manoeuvre and a clutter is generated around by coincidence, the clutter which turns out to more likely from the target than the real measurement can easily be confused with the real measurement. To address this dilemma which challenges the traditional filter too, % as the topic of our future work,
a potential solution is to carry out the density-based clustering scheme %\citep{Li17clustering,Li17MSDC,Li18FtC}
on the time-series measurements $\{\mathbf{Y}_i\}_{i=k-T_\text{s}+1}^k$ over a sliding time window from $k-T_\text{s}+1$ to the current time $k$ for identifying the potentially real measurements in the time window. The clustered measurements can then be used for fitting a 2-D spatial trajectory in the format of $y=f'(x)$.
%%%where, different from T-FoT, no time-variable is involved.
Then, % following the NN rule,
the measurement in the random finite set $\mathbf{Y}_k$ that matches best $y=f'(x)$ will be identified as the potentially real measurement of the target at time $k$, instead of using \eqref{eq:One-step_Prediction}-\eqref{eq:gating_y_NN}. However, the price that has to be paid for this is a higher computation requirement.
}

\subsubsection{T-FoT updating}
Denote by set ${K}' \subseteq [k', k]$ all the time instants having real measurements, where $k' = \text{max}(1,k-T)$, $T$ is the length of the time-window. % while correspondingly, all the left time instants have missed detecting the target.
Based on these measurements ${\mathbf{y}}_i, i\in K'$% in the sliding time-window $K_k$ up to the curret time $k$
, the parameters of T-FoT is now ready to be updated as shown in \eqref{eq:Exact_solution_for_Ck} for $d=1,2$, respectively. %The covariance of the estimate of $\mathbf{C}_k$ is the same as given in \eqref{eq:covC}. %\eqref{eq:Ck_optimization} or \eqref{eq:Ck_optimization}.
% \begin{equation} \label{eq:FoT_updating}
% \hat{\mathbf{C}}_{K} = \underset{\mathbf{C}_k}{\text{argmin}}\sum_{t\in K_k}
% \mathcal{D}_i(\mathbf{C}_k)
% %\big(\mathbf{y}_i-h_i({F}(t;\mathbf{C}_k),\bar{\mathbf{v}}_i)\big)^{\text T} \Sigma_{\mathbf{y}_i}^{-1}\big(\mathbf{y}_i-h_i({F}(t;\mathbf{C}_k),\bar{\mathbf{v}}_i)\big) .
% \end{equation}

%The resulting T-FoT parameters $\hat{\mathbf{C}}_{[k', k]}$  will be used for calculating a ``better'' state estimate conditioned on the newest measurement as compared with the predicted estimate \eqref{eq:One-step_Prediction}, and be used to iterate the above four steps for T-FoT updating at scanning time $k+1$.%, i.e.,
% \begin{equation} \label{eq:FoT-k}
% \hat{\mathbf{x}}_k={F}(t;\hat{\mathbf{C}}_{k}) .
% \end{equation}

\subsection{Trajectory Termination and Potential Re-start}
If the number of successive misdetections, $m$, is larger than $T_\text{e}$, target death will be declared and the trajectory should be terminated. %It has similar meaning of latency as $T_\text{s}$ does and so their values can be the same. % We set $T_\text{e}=T_\text{s}$ in our simulation.
The target may re-appear in the area and so the algorithm needs to re-do target detection (and then tracking) after terminating one T-FoT. This is also helpful when the algorithm terminates the trajectory wrongly due to more than $T_\text{e}$ successive misdetections. Since our proposed approach does not assume the target birth model, the target re-birth model is not needed.
If the target disappears shortly, it would be helpful to initialize the clustering procedure around the area where the target disappears. This can speed up the clustering calculation and lead to better accuracy. However, to ensure the best generality, we do not make such an assumption.
\SPrevis{We do not consider simultaneously detecting other targets while tracking one since it is assumed that there is no more than one target in the scenario. We leave the more challenging multiple target tracking issues to the future study.}

\begin{figure}
\centering
\includegraphics[width=10 cm]{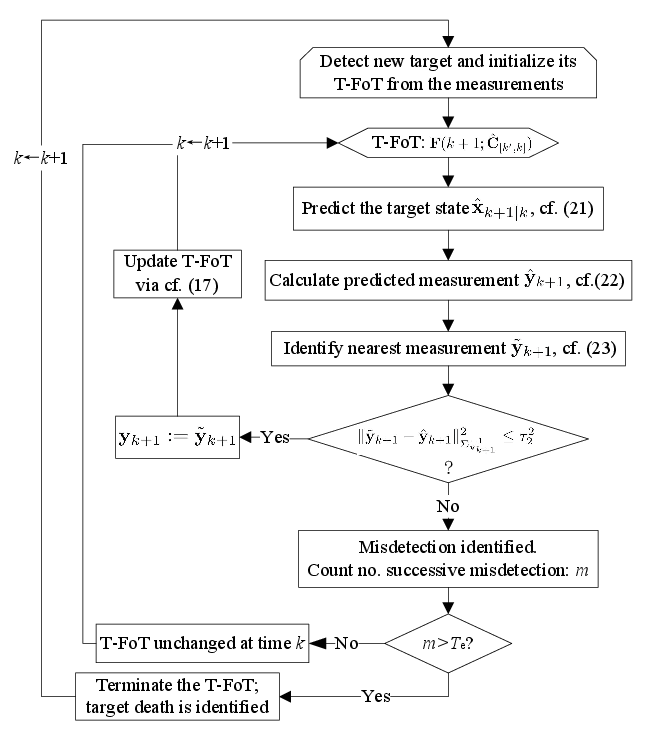}
\vspace{-2mm}
\caption{Procedure of the proposed JDT approach }\label{fig:diagram}
\end{figure}

\subsection{Overview of All Parameters Needed}
The whole procedure of the proposed JDT approach is given in Fig~\ref{fig:diagram}. %, which is subject to the constraint that there is no more than one target in the scenario.
In addition to the measurement function $h_k(\cdot)$ and the measurement noise statistics $\Sigma^{(d)}_{\mathbf{v}_i}$ which are needed in our algorithm, we summarize all the parameters needed in our approach in Table \ref{tab:1} as well as the favorable value for each parameter used in our simulation.
\begin{itemize}
  \item Both $\tau_1$ and $\tau_2$ are Mahalanobis distance-based and reasonably related to the uncertainty of the measurements, namely the magnitude of the measurement noise.
  \item $\gamma$ depends on the ``smoothness'' of the trajectory \citep{Li19stf} for which $\gamma=1, 2$ suit CV and CA targets, respectively.
  \item $T, T_\text{s}, T_\text{e}$ are determined more or less in a heuristic manner. As a rule of thumb, we specify $T_\text{s}, T_\text{e}$ in the scope of $[3,5]$ and $T$ in $[8, 12]$.
\end{itemize}

We note that in almost all trackers, confidence-based thresholds (relative to, e.g., target existing probability in the BF \citep{Vo12Bernoulli,Ristic13Bernoulli}; see also \citep[Ch.2]{Poor94} and \citep{Leung96,Tartakovsky20,Li21initialization}) are needed, %Ji19Initialization,
whether explicitly or implicitly, to initialize and terminate the target track. This indicates the heuristic nature of $T_\text{s}$ and $T_\text{e}$, \newrev{as well as the time-window length $T=10$, which depend on the practitioner}. As a rule of thumb, $T=10$ sensing steps turn out to be a close-to-best choice for the length of the sliding time-window in most of the scenarios we have tested.

\begin{table}[t!]
\renewcommand{\baselinestretch}{1.17}\small
\caption{Parameters for Our Proposed Approach}
\vspace{-4mm}
\label{tab:1}
\begin{center}
{\footnotesize
\begin{tabular}{|c|c|c|}
\hline
%\rule[-1.3mm]{0mm}{4.2mm}
Parameters & Purpose/meaning & Value Used \\
\hline
%\rule[-1.3mm]{0mm}{4.2mm}
$T_\text{s}$	&least latency for target detection & 4 \\
\hline
$T_\text{e}$	&least latency for confirming target death  & 4 \\
\hline
%\rule[-1.3mm]{0mm}{4.2mm}
$\tau_1$  & used in clustering for target detection & 3 \\
\hline
$\tau_2$	&used for identifying missing detection & 5 \\
\hline
$\gamma$	&order of the fitting function & 1  \\
\hline
$T$	&length of the time window for fitting & 10 \\
\hline
\end{tabular}
}
\end{center}
\vspace{-2mm}
\end{table}

\section{Simulations} \label{sec:simulation} %\revis{-----the whole section is new}

This section evaluates the performance of our approach and compare it with the Bayes-optimal BF \citep{Vo12Bernoulli,Ristic13Bernoulli} in two different scenarios. One uses a CV target model and a position measurement model while the other uses a coordinated turn (CT) target model and a range-bearing measurement model. They are referred to as linear and nonlinear systems, respectively. In each system, different parameters are used for generating different ground truths and scenarios. The simulation for each scenario is performed for 1000 Monte Carlo runs, each run using different target-trajectories and measurement series, all randomly generated based on the specified statistical models and parameters. \SPrevis{As usual, we still use the SSM for describing the kinematic model of the target. This is necessary for running a filter but not for our approach. }
While our approach uses no a-priori information about the target, the comparison BFs are provided with perfect a-priori information about the birth, death, and dynamics of the target, for their best possible performance. That is, the target is non-cooperative to our approach but cooperative to the BFs.
This gives an edge to the latter.

The performance of the filter and the proposed T-FoT approach is evaluated by the optimal subpattern assignment error (OSPA)
of the position estimation with cut-off $c = 1000$m and order $\rho \!=\! 2$ \citep{Schuhmacher08}. The OSPA between two RFSs $\hat{\mathbf{X}}$ and $\mathbf{X}$ is defined as follows, for $|\hat{\mathbf{X}}| \geq |\mathbf{X}|$,
\begin{equation} \label{eq:ospa}
d^{(c,p)}_\text{ospa}(\hat{\mathbf{X}}, \mathbf{X}) = \left( \frac{1}{|\hat{\mathbf{X}}|}\left( d_\text{Loc}(\hat{\mathbf{X}}, \mathbf{X})   +  d_\text{Card} (\hat{\mathbf{X}}, \mathbf{X}) \right) \right)^{\frac{1}{p}}
\end{equation}
which consists of two components accounting for the localization error and cardinality error, respectively, i.e.,
$d_\text{Loc}(\hat{\mathbf{X}}, \mathbf{X})  = {\mathop {\min }\limits_{\pi  \in {{\rm \Pi} _{|\hat{\mathbf{X}}|}}} \sum\limits_{i = 1}^{|\mathbf{X}|} {{d^{(c)}}{{({\mathbf{x}_i},{\mathbf{\hat{x}}_{\pi (i)}})}^p}} } $,
$d_\text{Card} (\hat{\mathbf{X}}, \mathbf{X})  = { {{c^p}}  (|\hat{\mathbf{X}}| - |\mathbf{X}|)}$.
%\begin{align}
%d_\text{Loc}(\hat{\mathbf{X}}, \mathbf{X}) & = {\mathop {\min }\limits_{\pi  \in {{\rm \Pi} _{|\hat{\mathbf{X}}|}}} \sum\limits_{i = 1}^{|\mathbf{X}|} {{d^{(c)}}{{({\mathbf{x}_i},{\mathbf{\hat{x}}_{\pi (i)}})}^p}} } , \\
%d_\text{Card} (\hat{\mathbf{X}}, \mathbf{X}) & = { {{c^p}}  (|\hat{\mathbf{X}}| - |\mathbf{X}|)}.
%\end{align}
Here, $\pi$ and $ {\rm \Pi}_n $ are a permutation and the set of all permutations on $\{1,\ldots,n \}$, and $ {d^{(c)}}(\mathbf{x},\mathbf{y}) = \min \left( {d(\mathbf{x},\mathbf{y}),c} \right) $ is a metric between $\mathbf{x}$ and $\mathbf{y}$ cut-off at $c$. If $|\hat{\mathbf{X}}| < |\mathbf{X}|$, $d^{(c,p)}_\text{ospa}(\hat{\mathbf{X}}, \mathbf{X}) = d^{(c,p)}_\text{ospa}(\mathbf{X}, \hat{\mathbf{X}})$. %, $d_\text{Loc}(\hat{\mathbf{X}}, \mathbf{X}) = d_\text{Loc}(\mathbf{X},\hat{\mathbf{X}})$, $d_\text{Card} (\hat{\mathbf{X}}, \mathbf{X})  = d_\text{Card} (\mathbf{X}, \hat{\mathbf{X}}) $.

%Our simulations are based on MATLAB (R2018a) implementations on an Intel Core i7--8750H CPU. \SPrevis{The source codes and simulation data are available at\\ https://drive.google.com/file/d/1MwEqoYzABBoBUiXSSSafQufQFDnF2SgP/view?usp=sharing and via emails to the First Author}.

\subsection{Linear System}
In this simulation lasting for $100$s, both the dynamic model and the measurement model are linear. Denote the target state as $\mathbf{x}_k =[p_{k,x}, \dot{p}_{k,x}, p_{k,y}, \dot{p}_{k,y}]^{\mathrm{T}}$, which is composed of
%% 2-D
position $[p_{x,k},p_{y,k}]^\mathrm{T}$ and
%% 2-D
velocity $[\dot{p}_{x,k},\dot{p}_{y,k}]^\mathrm{T}$. %, i.e., $\mathbf{x}_k={F}\mathbf{x}_{k-1}+\mathbf{G}\mathbf{u}_{k}$.
The target birth Bernoulli intensity function is given as
\begin{equation} \label{eq:new_birth1}
\gamma_k(\mathbf{x}_k)  =  0.01 \mathcal{N}(\mathbf{x}_k; \mathbf{m}, \mathbf{P}),
\end{equation}
where $\mathbf{m} \!=  [-500 \text{m}, 10 \text{m}/\text{s}, -500 \text{m}, 10 \text{m}/\text{s}]^\text{T}$,
$\mathbf{P} = \mathrm{diag} (100 \text{m}, 10 \text{m}\!/\text{s}, 100 \text{m}, 10 \text{m}\!/\text{s})^2$ .
%% , where diag$(\mathbf{a})$ representing a diagonal matrix with elements along the diagonal given by $\mathbf{a}$.

The target appears at time $k=10$s with a random initial
state according to the above newborn target model and disappears at time $k=80$s. To model this, the target survival probability is set as ${0.99}$.
%%  with position $[p_{x, k}, p_{y, k}]^\mathrm{T}$ and velocity $[\dot{p}_{x, k}, \dot{p}_{y, k}]^\mathrm{T}$
During time $k\in [10\text{s}, 80\text{s}]$, the state of the target evolve according to a nearly CV model, i.e.,
$\mathbf{x}_k= {F} \mathbf{x}_{k-1} + \mathbf{G}\mathbf{u}_{k-1}$,
with
$$ {F} = \mathbf{I}_2\otimes \left[ \begin{array}{cc}
1 & \Delta \\
0 & 1 \\
\end{array} \right],
\mathbf{G} = \left[ \begin{array}{l}
\mathbf{I}_2 \otimes \frac{\Delta^2}{2} \\
\mathbf{I}_2 \otimes \Delta \\
\end{array} \right], $$
where $\Delta \!=\! 1$s and $\mathbf{u}_{k-1}$ is zero-mean Gaussian noise with time-invariant covariance $\mathbf{Q}$ of unit $\textrm{m}^2/\textrm{s}^4$.
% \mathcal{N}(\mathbf{u};\mathbf{0}_2\textrm{m}/\textrm{s}^2,\delta^2\mathbf{I}_2\textrm{m}^2/\textrm{s}^4)$.

The target detection probability denoted by $p^\text{D}$ is constant. Once the target is detected, it generates a position measurement as in \eqref{eq:Position_Measure} with $\mathbf{v}_{k}$ being a zero-mean Gaussian noise with covariance $\Sigma_{\mathbf{v}_{k}} = \textrm{diag} (100\hspace{.5mm}\mathrm{m}^2, 100\hspace{.5mm} \mathrm{m}^2)$. The clutter is uniformly distributed over the planar area $[-1000\text{m},1000\text{m}]\times [-1000\text{m},1000\text{m}]$ with an average of $r_c$ points per scan. We note here that, the target may fly out of the mentioned area in some runs. The clutter density is $r_c/2000^2$m$^{-2}$ in the cluttered area $[-1000\text{m},1000\text{m}]\times [-1000\text{m},1000\text{m}]$ and zero outside. However, the BF assumes a constant clutter intensity at $r_c/2000^2$m$^{-2}$. % \footnote{We have tested setting the clutter intensity needed by the filter outside the cluttered area as zero and using a smaller target velocity so that the target is always inside the cluttered area, respectively. However, the results make little change and the comparison results hold the same. }.
% for generating the target trajectories (see Fig.~\ref{fig:network}).
In our simulation, we test different values for $p^\text{D}$, $r_c$, and $\mathbf{Q}$.

Our approach uses the first order polynomial %. fitting function in \eqref{eq:Ck_optimization} The optimization \eqref{eq:Ck_optimization} is carried our
%. That is, the polynomial
T-FoT in $x-$ and $y-$ dimensions, respectively, as follows
\begin{align}
p_{x,t} & = c_0^{(1)} + c_1^{(1)} t + \mathbf{\epsilon}^{(1)}(t), \label{eq:sim1_T_FoT1}\\
p_{y,t} & = c_0^{(2)} + c_1^{(2)} t + \mathbf{\epsilon}^{(2)}(t), \label{eq:sim1_T_FoT2}
\end{align}
where the parameters $\mathbf{C}_{K'}:=\{c_0^{(1)},c_1^{(1)},c_0^{(2)},c_1^{(2)}\}$ are calculated by carrying out the LS optimization as in \eqref{eq:Ck_optimization} over the time window $K' \subseteq [k', k]$, for $d=1,2$, respectively. Here, $k' = \text{max}(1,k-T)$, $T=10$s is the length of the time-window, and $K'$ contains all the time-instants when there is a target detection in $[k', k]$.

%It is reasonable to assume that the statistics of both $\mathbf{\epsilon}^{(1)}(t)$ and $\mathbf{\epsilon}^{(2)}(t)$ are actually time-invariant (i.e., the fitting quality is the same on average at all times over the time-window). This,
Applying \eqref{eq:approx_e}
with the time-invariant statistics of the measurement noise $\mathbf{v}_k$ leads to the time-invariant statistics of the fitting error $\mathbf{e}_t$. So, $\Sigma_{\mathbf{e}_t}$ can be removed from \eqref{eq:Ck_optimization}, i.e., %. That is, when the statistics of the measurement is known to be constant over time and unrelated to the state of the target, the weighted LS estimation reduces to the ordinary LS estimation as follows: % can be interpreted as the measurement noise similar to in \eqref{eq:MeasurementModel}, while ${F}(t;\mathbf{C}_k)$, $h_t(\cdot,\bar{\mathbf{v}}_t)$ being the state and measurement function, respectively. As such, we use $\Sigma_{\mathbf{y}_t} = \Sigma_{\mathbf{e}_t}$.
% \mathbf{e}_t $ $\Sigma_{\mathbf{y}_t}$ is the covariance of the measurement. %The LS error is our focus in this work. %, which is equivalent to that of the noise in the case of additive noise.
%  In the case of time-invariant measurement noise statistics, %, which does not need to know the covariance of the noise exactly.
\begin{equation}
  \big\{ c_0^{(d)},c_1^{(d)} \big\}  = \underset{\{ c_0,c_1\}}{\text{argmin}} \sum_{t \in K'} \left(y^{(d)}_{t}-(c_1 t + c_0 )\right)^2 , \label{eq:linearFoT_optimization}
\end{equation}
%where
%\begin{equation} \label{eq:linearFoT_optimization}
%\hat{f}(t) =
%\underset{{F}(t;\mathbf{C}_k)}{\text{argmin}} \sum_{i=k-T}^{k} (\mathbf{y}_i-\hat{\mathbf{y}}_i)^{\text T} (\mathbf{y}_i-\hat{\mathbf{y}}_i),
%\end{equation}

Our approach does not use any mentioned target birth, death and dynamics, or clutter rate information but only need to use $\Sigma_{\mathbf{v}_{k}}$ in the initial clustering operation as shown in \eqref{eq:tau_1} for detecting the target. The values of all parameters needed are given in Table~\ref{tab:1}. The estimated position of the target at time $k$ is simply given by substituting $t$ by $k$ in \eqref{eq:sim1_T_FoT1} and \eqref{eq:sim1_T_FoT2}. %As addressed in \eqref{eq:Exact_solution_for_Ck} (using $\Sigma_d = \mathbf{I}_{|K'|}$), \eqref{eq:linearFoT_optimization} can be solved efficiently. %At present, we do not further consider velocity estimation.

As the comparison approach, a Gaussian mixture (GM) BF \citep{Vo12Bernoulli,Ristic13Bernoulli} is designed using all the mentioned necessary statistics information about the target, clutter, and the sensor. %This ensures the best possible performance of the BF.
Furthermore, the BF used at most 50 Gaussian components (GCs), pruned GCs with weights below $10^{-5}$ and merged GCs with Mahalanobis distance below $4$.

\begin{figure}
\centering
\includegraphics[width=9 cm]{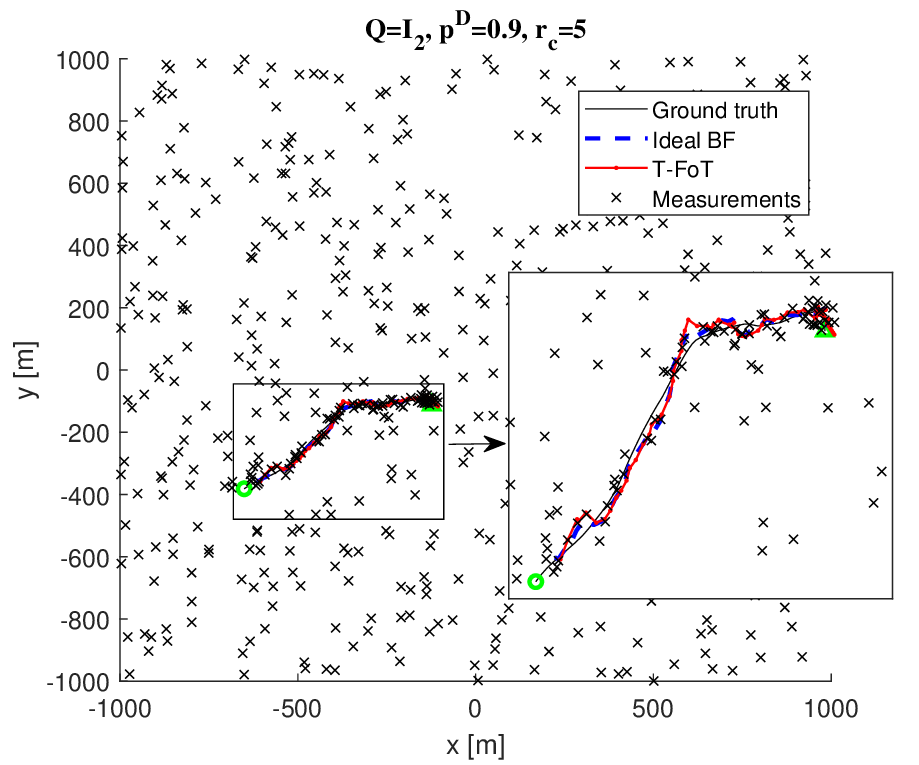} %6.5
\vspace{-2mm}
\caption{Target trajectory and measurements generated over 100 seconds in the linear system using process noise covariance $\mathbf{Q}=\mathbf{I}_2$, target detection probability $p^\text{D}=0.9$ and clutter rate $r_c=5$ in one run. The green circle and triangle indicate the start and end of the trajectory, respectively. }\label{fig:LinearScen_Q1}
\vspace{-2.5mm}
\end{figure}

\begin{figure}
\centering
\includegraphics[width=10 cm]{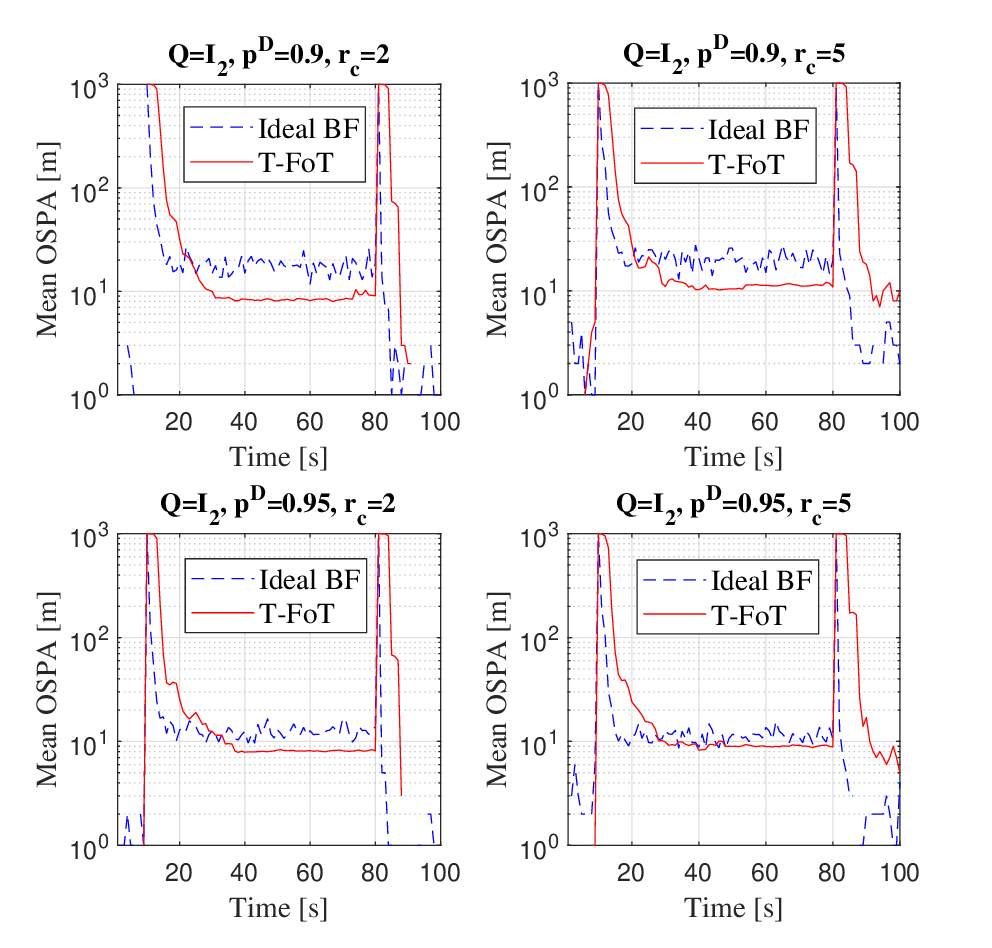}
\vspace{-2mm}
\caption{OSPA of the BF and our T-FoT approach in the linear system using $\mathbf{Q}=\mathbf{I}_2$, and different target detection probabilities $p^\text{D}$ and clutter rates $r_c$. }\label{fig:Linear_Q1}
\vspace{-2.5mm}
\end{figure}

\begin{figure}
\centering
\includegraphics[width=9 cm]{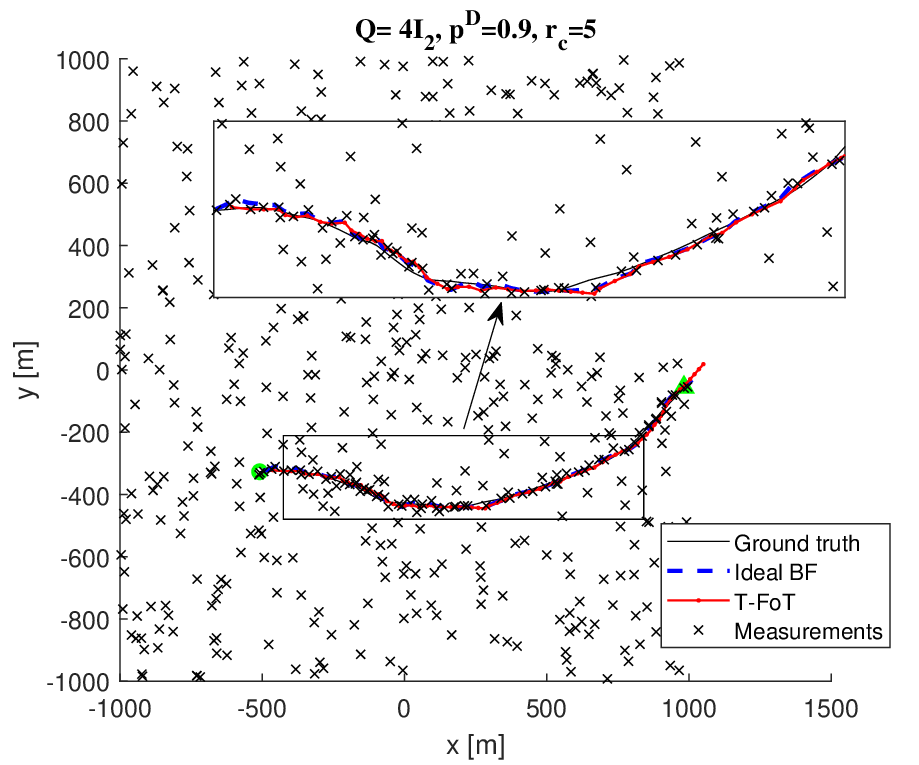}
\vspace{-2mm}
\caption{Target trajectory and measurements generated over 100 seconds in the linear system using process noise covariance $\mathbf{Q}=4\times\mathbf{I}_2$, target detection probability $p^\text{D}=0.9$ and clutter rate $r_c=5$ in one run. The green circle and triangle indicate the start and end of the trajectory, respectively. }\label{fig:LinearScen_Q4}
\vspace{-2.5mm}
\end{figure}

\begin{figure}
\centering
\includegraphics[width=10 cm]{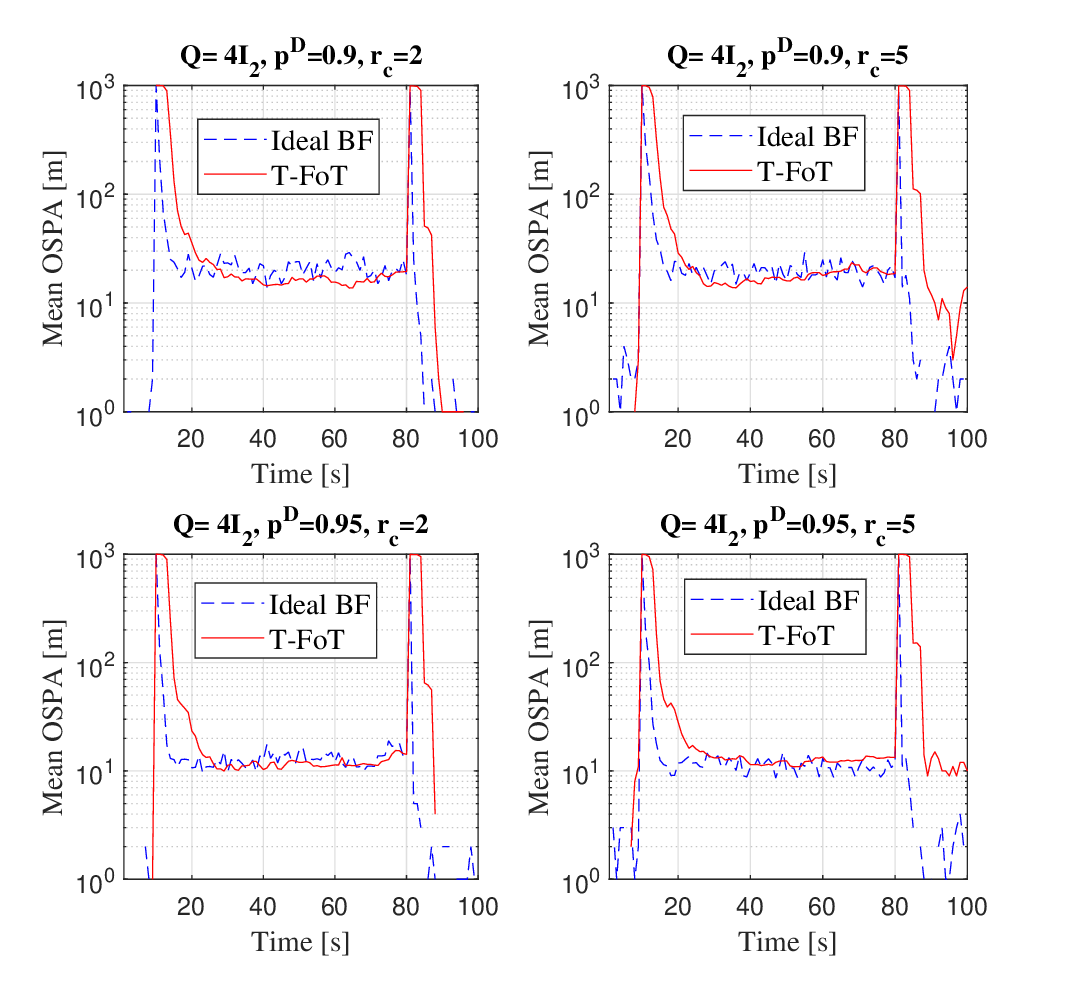}
\vspace{-2mm}
\caption{OSPA of the BF and our T-FoT approach in the linear system using $\mathbf{Q}=4\times\mathbf{I}_2$, and different detection probabilities $p^\text{D}$ and clutter rates $r_c$.}\label{fig:Linear_Q4}
\vspace{-2.5mm}
\end{figure}

The average results of 1000 Monte Carlo runs are shown in Fig.~\ref{fig:Linear_Q1} and Fig.~\ref{fig:Linear_Q4} for process noise covariance $\mathbf{Q}=\mathbf{I}_2$ and $\mathbf{Q}=4\times\mathbf{I}_2$, respectively. In each figure, the subfigures show the results for different target detection probabilities $p^\text{D}=0.9, 0.95$ and different clutter rates $r_c=2,5$. The tracking scenario of one run using $p^\text{D}=0.95$ and $r_c=5$ is illustrated in Fig~\ref{fig:LinearScen_Q1} for $\mathbf{Q}=\mathbf{I}_2$ and in Fig~\ref{fig:LinearScen_Q4} for $\mathbf{Q}=4\times\mathbf{I}_2$. The computing time of each updating step for all scenarios is given in Table~\ref{tab:sim1}. These results show that our proposed T-FoT approach computes much faster, yields higher track accuracy (when the target exists) but is less accurate in identifying the birth and death of the target as compared with the optimal BF. In particular, we notice that
\begin{enumerate}
\item The T-FoT approach suffers from target detection and track termination latency more than the BF as its OSPA reduces down to a stable low level slower than the BF does. This is mainly due to the lack of information about target birth and death. To combat this, one may reduce $\tau_1$ for faster track initiation and reduce $T_e$ for faster track termination. These, however, come with the price of causing false alarms and with the risk of premature killing the track. How to better trade-off both, which depends on the realistic need, remains an open problem.
  \item For all cases except for $\mathbf{Q}=4\times\mathbf{I}_2, p^\text{D}=0.95,r_c=5$, our proposed T-FoT approach outperforms the BF in mean OSPA during time $k \in [30\text{s},80\text{s}]$. This may be attributed to two reasons: First, the GM implementation of the BF that unavoidably bears approximation errors prevents the BF from achieving the theoretically best performance. % (e.g., Bayesian Cramer-Rao lower bound, BCRLB),
       \newrev{To say the least, its Bayes-optimization does not equal minimum OSPA as the latter is no more than a specific yet common point-based metric.} Second, our proposed approach is based on sliding time-window fitting which is not as sensitive to clutter or misdetection at any particular time-instant as the filter. % So, a filter will be more severely affected by whether clutter or misdetection than in our approach. % which does not impose first-order Markov assumption like the filter. % and is not necessarily bounded by the BCRLB. %is more like a smoothing approach rather than a filtering approach. %
      %First, although the BF is Bayesian optimal, its implementation based on a GM bears unavoidably approximation errors. % Second, the time-invariant clutter intensity assumed by the BF does match exactly the case when the target moves out of the cluttered area -- this, however, . %This is somehow surprising at the first glace.
      For the scene using $\mathbf{Q}=4\times\mathbf{I}_2, p^\text{D}=0.95,r_c=5$, both perform similar to each other. But in all cases, the T-FoT approach has an average OSPA varying over time in a smaller scope when the target exists and is tracked during $k \in [30\text{s},80\text{s}]$.
  \item It is obvious that the performance of the proposed T-FoT approach is much better when $\mathbf{Q}= \mathbf{I}_2$ (which corresponds to a smoother trajectory) than when $\mathbf{Q}=4\times\mathbf{I}_2$. In contrast, the performance of the BFs does not change much with the change of $\mathbf{Q}$ or $r_c$, but it becomes much better if the target detection probability is increased from $p^\text{D}=0.9$ to $p^\text{D}=0.95$.
  \item A higher clutter rate renders our approach prone to declaring appearance of the target (e.g., at time $k\in[85\text{s}, 100 \text{s}]$). This is reasonable as it used no a-priori information about the target birth and so the algorithm can easily be misled by the locally clustered clutter in successive scans when the clutter rate is high. In contrast, the BF has a proper target birth model to prevent it from wrongly initializing a new track that does not match closely to the specified target-birth model.
\end{enumerate}

\begin{table}[t!]
\renewcommand{\baselinestretch}{1.17}\small
\caption{Computing Time for Each Step in Simulation 1 [second]}
\vspace{-4mm}
\label{tab:sim1}
\begin{center}
{\footnotesize
\begin{tabular}{|c|c|c|}
\hline
%\rule[-1.3mm]{0mm}{4.2mm}
Scenario & BF & T-FoT \\
\hline
%\rule[-1.3mm]{0mm}{4.2mm}
$\mathbf{Q}=\mathbf{I}_2,p^\text{D}=0.9, r_c=2$	&  0.041  & 0.016 \\
\hline
$\mathbf{Q}=\mathbf{I}_2,p^\text{D}=0.9, r_c=5$	&  0.049  & 0.017 \\
\hline
%\rule[-1.3mm]{0mm}{4.2mm}
$\mathbf{Q}=\mathbf{I}_2,p^\text{D}=0.95, r_c=2$	&  0.039  & 0.016 \\
\hline
$\mathbf{Q}=\mathbf{I}_2,p^\text{D}=0.95, r_c=5$	&  0.047  & 0.017 \\
\hline
$\mathbf{Q}=4\mathbf{I}_2,p^\text{D}=0.9, r_c=2$	&  0.043  & 0.017 \\
\hline
$\mathbf{Q}=4\mathbf{I}_2,p^\text{D}=0.9, r_c=5$	&  0.051  & 0.017 \\
\hline
$\mathbf{Q}=4\mathbf{I}_2,p^\text{D}=0.95, r_c=2$	&  0.041  & 0.016 \\
\hline
$\mathbf{Q}=4\mathbf{I}_2,p^\text{D}=0.95, r_c=5$	&  0.048  & 0.017 \\
\hline
\end{tabular}
}
\end{center}
\vspace{-2mm}
\end{table}

\vspace{-3mm}
\subsection{Nonlinear System}
\label{sec:nonli-Sim}
In this simulation of a length of $150$s, we consider an CT target motion model and a range-bearing measurement model. The target state is denoted as $\mathbf{x}_k=[p_{x,k} \; \dot{p}_{x,k} \; p_{y,k} \; \dot{p}_{x,k} \;\omega _{k}]^\text{T} $ with time-varying turn rate $\omega _{k}$.
The target birth intensity function is given as
\begin{equation} \label{eq:new_birth2}
\gamma_k(\mathbf{x}_k)  =  0.01 \times \mathcal{N}(\mathbf{x}_k; \mathbf{m}_1, \mathbf{P}_1),
\end{equation}
where $\mathbf{m}_1=[100 \text{m}, 10 \text{m}/\text{s}, 100 \text{m}, 10 \text{m}/\text{s},0.01\text{rad}]^\text{T}$ and $\mathbf{P}_1 \!=  \mathrm{diag} (100 \text{m}, 10 \text{m}/\text{s}, 100 \text{m}, 10 \text{m}/\text{s}, 0.01\text{rad})$.

One target appears at time $k=10$s with a random initial state according to the above newborn target model and disappears at time $k=80$s. The probability of target survival is $P^\text{S}_k=0.99$ for the BF. The survival single-target movement follows a CT model with a
sampling period of $1$s and Markov transition function
\begin{equation}\label{eq:nonlinearMarkovSim2}
  f_{k|k-1}(\mathbf{x}_{k}|\mathbf{x}_{k-1})=%
\mathcal{N}(\mathbf{x}_{k};F(\omega _{k})\mathbf{x}_{k},\mathbf{Q}) ,
\end{equation}
where $\mathbf{Q}=\mathrm{diag}([\mathbf{I}_2\otimes\mathbf{G},\sigma _{u}^{2}])$,
$$ F(\omega )=\left[
\begin{array}{ccccc}
1 & \!\!\frac{\sin \omega }{\omega } & 0 & \!\!-\frac{1-\cos \omega }{\omega
} & 0 \\
\vspace{0.5mm}
0 & \!\!\cos \omega & 0 & \!\!-\sin \omega & 0 \\
\vspace{0.5mm}
0 & \!\!\frac{1-\cos \omega }{\omega } & 1 & \!\!\frac{\sin \omega }{\omega }
& 0 \\
\vspace{0.5mm}
0 & \!\!\sin \omega & 0 & \!\!\cos \omega & 0 \\
\vspace{0.5mm}
0 & 0 & 0 & 0 & 1%
\end{array}%
\right] \!\!, \\
\mathbf{G}=%
\left[
\begin{array}{cc}
\vspace{0.8mm}
\frac{\sigma _{w}^{2}}{4} & \frac{\sigma _{w}^{2}}{2} \\
\vspace{0.8mm}
\frac{\sigma _{w}^{2}}{2} & \sigma _{w}^{2}  %
\end{array}%
\right] \!\!
$$
with $\sigma_{w}=2 \text{m}/\text{s}^{2}$, and $\sigma _{u}=(\pi /180)\text{rad}/\text{s}$.

Further, we consider the re-appearance of the target. It is simulated that the target re-appears at time $k=90$s and disappears again at time $k=110$s. (Sure, this can be viewed as another new target). The initial state $\mathbf{x}_{90}$ is given by $\mathbf{x}_{90}  \sim \mathcal{N}(\mathbf{x}_k; \mathbf{m}_2, \mathbf{P}_2)$,
where $\mathbf{m}_2=[500 \text{m}, 10 \text{m}/\text{s}, 500 \text{m}, 10 \text{m}/\text{s},0.01\text{rad}]^\text{T}$ and $\mathbf{P}_2 \!=  \mathrm{diag} (100 \text{m}, 10 \text{m}/\text{s}, 100 \text{m}, 10 \text{m}/\text{s}, 0.01\text{rad})$.

%We considered the following observation model.
The range-bearing measurement model is given as in \eqref{eq:range-bearing-measurement} where %for which the sensor located at the coordinate origin,
$v_{{r},k} $ and $v_{\theta,k} $ are, individually, independent identical distributed zero-mean Gaussian with standard deviation $\sigma_{r} \!=\! 10 $m and $\sigma_\theta =  (\pi/90) $rad, respectively.

The target detection probability is state-related as given by
%% \vspace{-1mm}
\begin{equation} \label{eq:sim2_state-related_D}
p_k^\text{D}(\mathbf{x}_k) = p_\text{max}^\text{D} \cdot \frac{ \mathcal{N}\big( \mathbf{\mu}_\mathrm{D}(\mathbf{x}_k); \mathbf{0},2000^2 \mathbf{I}_2\big)}{
  \mathcal{N}(\mathbf{0};\mathbf{0},2000^2\mathbf{I}_2)} .
\vspace{-.5mm}
\end{equation}
Here, %$p^\text{D} \rmv=\rmv 0.95$, % or $0.8$and
$\mathbf{\mu}_\mathrm{D}(\mathbf{x}_k) \triangleq  \big[ | p_{x,k} |, | p_{y,k}| \big]^\text{T}\!$.
%% $[ x^{(s)} \; y^{(s)}]^\text{T}$ is the position of sensor $s$.

The clutter measurements are uniformly distributed over a  disk
of radius $2000$m around the origin with an average number of $r_\text{c}$ clutter measurements per time step. %, or equivalently clutter intensity $ r_\text{c}/(\pi  \cdot  2000)$.
We considered two different clutter rates $r_c=2,5$ and two different maximal detection probabilities $p_\text{max}^\text{D}=0.9, 0.95$.

In our approach for T-FoT fitting, the range-bearing measurements $\mathbf{y}_{k}$ are converted to position measurements as in \eqref{eq:rang-bearing-conversion} and the converted noise covariances are calculated based on linearization \citep{Bordonaro14}. The only a-priori statistical information used in our approach is about $\sigma_{r} $ and $\sigma_\theta$. Then, both clustering and fitting can be performed over the converted position measurement as in the last simulation, cf. \eqref{eq:sim1_T_FoT1} and \eqref{eq:sim1_T_FoT2}. However, we note that, differently from the last simulation, the converted covariance $\Sigma_{\mathbf{y}_{k}}$ is obviously state-related% and cross-correlated
, and therefore should be explicitly taken into account in the T-FoT optimization as shown in \eqref{eq:Ck_optimization}. % and consequently, the individual fitting result as given in \eqref{eq:Exact_solution_for_Ck} cannot be applied.
The multi-dimensional optimization, however, is computationally much more complicated (and did not yield obviously better results as we found). \SPrevis{Therefore, we over this difficulty by omitting the cross-correlation and carrying out individual dimension fitting.} %, which is exactly solved by \eqref{eq:Exact_solution_for_Ck}. That is, in place of \eqref{eq:linearFoT_optimization},
Furthermore, %can be interpreted as the measurement noise similar to in \eqref{eq:MeasurementModel}, while ${F}(t;\mathbf{C}_k)$, $h_t(\cdot,\bar{\mathbf{v}}_t)$ being the state and measurement function, respectively. As such,
we use $ \Sigma^{(d)}_{\mathbf{e}_i} \propto %= \mathrm{const} \times
\Sigma^{(d)}_{\mathbf{y}_i}$.
Then, we get the following weighted LS formulation
\begin{equation}
  \{ c_0^{(d)},c_1^{(d)} \}  = \underset{\{ c_0,c_1\}}{\text{argmin}} \sum_{i \in K'} \Sigma_{\mathbf{y}^{(d)}_{i}}^{-1}\big(y^{(d)}_{i}-(c_1i + c_0)\big)^2  . \label{eq:CovlinearFoT_optimization}
\end{equation}

For comparison, the local GM-BFs are implemented based on either the extended KF (EKF), unscented KF (UKF) or the particle filter (PF). In the former two cases, the filter used at most $50$ GCs, pruned GCs with weights below $10^{-5}$ and merged GCs with Mahalanobis distance below $4$ while in the latter, 2000 particles are allocated to the filter and 1000 particles are assigned to the target born process; see \citep{Vo12Bernoulli,Ristic13Bernoulli} for the detail of these algorithms. Again, these filters make full use of all available models and parameters.

\begin{figure}
\centering
\includegraphics[width=9 cm]{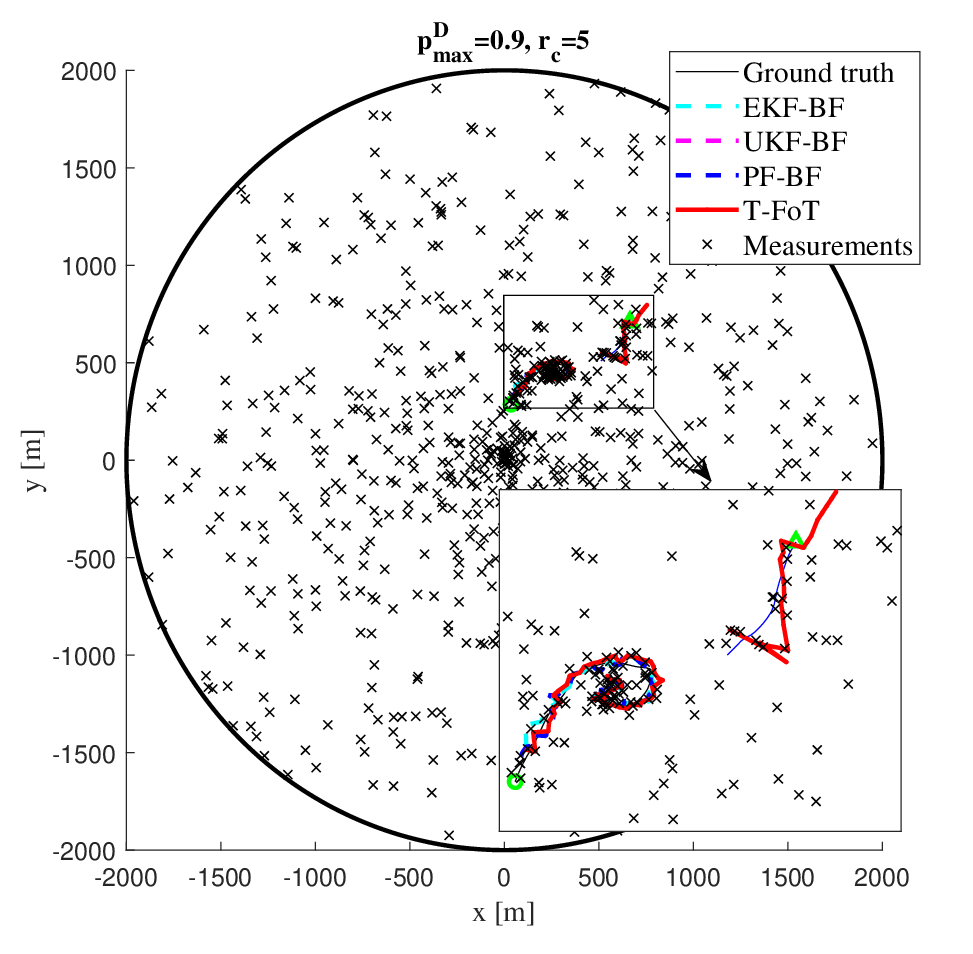}
\vspace{-2mm}
\caption{Target trajectory and measurements over 150 seconds generated by the nonlinear systemusing $p_\text{max}^D=0.9$ and $r_c=5$ in one run. Green circle and triangle indicate the start and end of the trajectory, respectively. }\label{fig:nonlinearScen}
\vspace{-2.5mm}
\end{figure}

\begin{figure}
\centering
\includegraphics[width=10 cm]{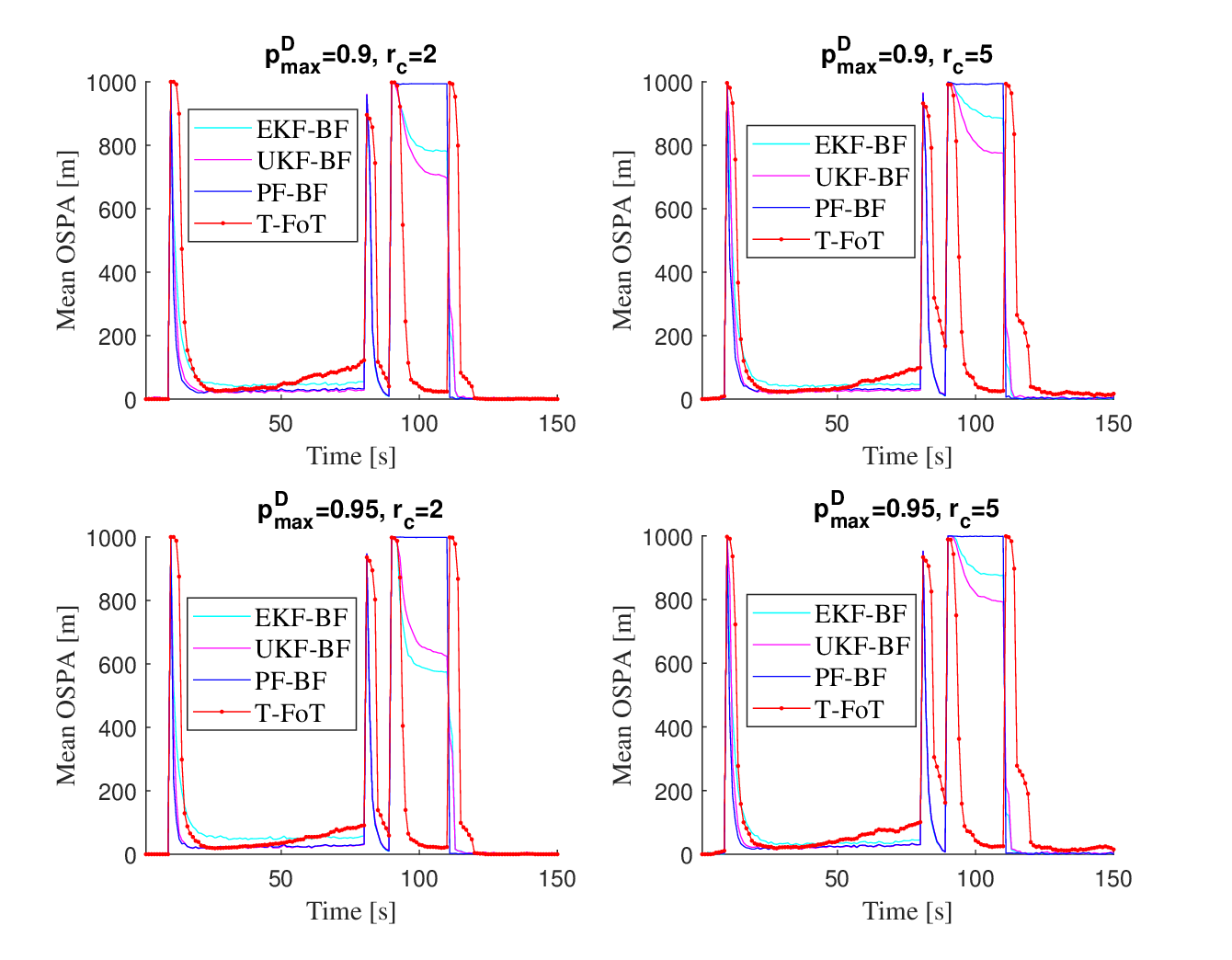}
\vspace{-2mm}
\caption{OSPA of BFs and our T-FoT approach in the nonlinear system using different target detection probabilities $p^\text{D}$ and clutter rates $r_c$.}\label{fig:nonlinearPerf}
\vspace{-2.5mm}
\end{figure}

The target trajectory and estimates by different filters and the proposed T-FoT approach in one run using $p^\text{D}=0.95$ and $r_c=5$ are illustrated in Fig~\ref{fig:nonlinearScen}.
The average OSPA of these comparison filters and the proposed T-FoT approach over 1000 Monte Carlo runs are given in Fig.~\ref{fig:nonlinearPerf}. The computing time of each updating step for all scenarios is given in Table~\ref{tab:sim2}. These results show that in this nonlinear system:
\begin{enumerate}
\item  The T-FoT approach suffers again from target detection and track termination latency more than the BFs, similar as in the first simulation. Also, the results after the disappearance of the second target $k > 120 \text{s}$ show that a higher clutter rate ($r_c=5$) can easier lead to false alarm to the T-FoT approach (as compared with $r_c=2$). The same reasons hold.
  \item The accuracy of the proposed T-FoT approach is comparable to the UKF/PF BFs (and outperforms the EKF-BF) during time $k \in [20\text{s}, 40\text{s}]$ but later on its accuracy decreases much faster than the BFs do. The accuracy decrease is due to the fact that in most runs, the target moves away from the origin and so the converted position accuracy of the range-bearing measurement as in \eqref{eq:range-bearing-measurement} is reducing
      and also the target detection probability is reducing as indicated by \eqref{eq:sim2_state-related_D}. \SPrevis{As can be easily illustrated, the position error corresponding to bearing error ``$\sigma_\theta$" at range distance of $100$m is almost 10 times smaller than that at a range distance of $1000$m.} %We also notice that in some runs, the target has a high turn rate which is actually unsuitable to be fitted by polynomial functions - there is space for online adjusting the fitting function adaptively.
      Moreover, \SPrevis{the converted position cross-correlation that we omitted in our fitting increases
      as the range increases}. This differs from the last simulation where the position measurements are of constant quality disregarding the state of the target and are not cross-correlated. % nonlinear conversion of the range-bearing measurements, which bears more or less conversion error. Whatever, divergence is indeed undesirable performance of the T-FoT approach for which at present we have no idea how to guarantee its stability and convergence.
  \item For the second, ``unexpected" target appearing at time $k=90$s, all BFs that do not set a corresponding model for it have missed it to a large degree while the T-FoT approach can quickly detect and then track it accurately the same as the first target. Arguably, both the success (i.e., quick detection) and failure (i.e., misdetection) of the filter depend on whether the specified models match perfectly the truth. %In contrast, our approach is data-driven and free of the model mismatch problem.
  \item The proposed T-FoT approach is computationally more efficient than all BFs. The T-FoT approach only takes slightly more computation due to the measurement conversion as compared with that in the first simulation, while the EKF/UKF BFs take almost two times computation time for each filtering step in order to deal with the nonlinearity in filtering. The computational cost of the PF-BF is even higher.
\end{enumerate}

\noindent

\begin{table}[t!]
\renewcommand{\baselinestretch}{1.17}\small
\caption{Computing Time for Each Step in Nonlinear System [second]}
\vspace{-4mm}
\label{tab:sim2}
\begin{center}
{\footnotesize
\begin{tabular}{|c|c|c|c|c|}
\hline
%\rule[-1.3mm]{0mm}{4.2mm}
Scenario & EKF-BF & UKF-BF& PF-BF & T-FoT \\
\hline
%\rule[-1.3mm]{0mm}{4.2mm}
$p^\text{D}=0.9, r_c=2$	&  0.094 &  0.112 &  0.170  & 0.036 \\
\hline
$p^\text{D}=0.9, r_c=5$	&  0.152 &  0.169 &  0.230  & 0.035 \\
\hline
%\rule[-1.3mm]{0mm}{4.2mm}
$p^\text{D}=0.95, r_c=2$	&  0.087 &  0.11 &  0.167  & 0.029 \\
\hline
$p^\text{D}=0.95, r_c=5$	&  0.138 &  0.160 &  0.226  & 0.029 \\
\hline
\end{tabular}
}
\end{center}
\vspace{-4mm}
\end{table}

 \vspace{-3mm}
\subsection{Discussion}
We must reiterate that the scenarios we considered here are limited to (near-)uniformly distributed clutter with a low rate ($r_c\leq5$) and high target detection probability ($p^\text{D}\geq0.9$). Relation of these limitations is valuable and deserves to be investigated.
As shown in both simulations, the proposed T-FoT approach is comparable with, or even better than, the BFs in computing efficiency and track maintenance accuracy although it suffers more from track confirmation and termination latency due to the lack of a-prior information about the target birth and death. What are additionally notable include
\begin{enumerate}
  \item %Although the proposed data-driven T-FoT approach has operations like density clustering and curve fitting over a sliding time window, it
  The T-FoT approach computes efficiently thanks to the exact solution given for linear fitting as shown in \eqref{eq:Exact_solution_for_Ck}.
  \item The fitting calculation over data of a time-window is resilience to the false/missing data at a particular time-instant as compared with the recursive filtering.
  \item The Bayes-optimization of the BF does not equal minimum OSPA. Obviously, the OSPA values depends on the choices of the parameters $c$ and $p$.
  \item \SPrevis{Although the BF is Bayes optimal, it generally has no analytic solution and has to resort to approximative implementation using either GM or particles. The number of GCs or particles used and the ineluctable merging/pruning/resampling operations can all have a significant effect on the filtering accuracy. These are attributed to the discrepancy of the realistic performance of the BF to the desired Bayes optimality.} %. The BF implementation needs to balance the approximation accuracy and computation efficiency. They are
\end{enumerate}
% for the apparently surprising result that our data-driven T-FoT approach outperforms the BFs in gaining better accuracy in scenarios, while using smaller computation cost. %, although it does not explicitly uses a prior statistical information about the target kinematic motion nor about the target detection/clutter ratios. %In our simulation, the ground truth was generated here by using Markov models, which resonates the dynamics of the recursive filter. If the ground truth is given by a curve T-FoT with frequent maneuvers but still smooth, the filter can hardly yield so good performance but our T-FoT approach may perform the same or even better.
%Finally, we emphasize that, the T-FoT-oriented estimator (i) needs no, but (ii) provides continuous.

%These being said, we have observed the tracking divergence of the T-FoT approach in the nonlinear systems. How to fix this remains unclear.
%More importantly, (iii) by T-FoT fitting, we do not have to worry about the target maneuver as long as no sharp maneuver occurs. This is significantly different from existing maneuvering target tracking approaches that rely on careful model design.

\section{Conclusion} \label{sec:conclusion}
We have presented a Markov-free, data-driven approach to JDT of a non-cooperative target that randomly appears and disappears in the presence of false and missing data. % while using little a-priori model information about target and the scenario.
While the clutter rate is less than 5 per scan on average and the target detection probability is higher than 90\%, their exact statistics, as well as those of the birth, death, and dynamics of the target are unknown and maybe time-varying. Our approach based on the T-FoT overcomes these challenges by only making use of measurements for joint target detection and continuous-time trajectory estimation including initiation, maintenance, termination and even re-detection. % which is more informative than a series of detection-points given by existing approaches.
Simulation results have demonstrated that our approach performs comparable to the properly modelled Bernoulli filter provided with all required model and scenario information and even outperforms them in some cases while computing more efficiently. Therefore, the T-FoT approach provides a promising alternative to the classic state space model and accommodates favorably intelligent learning methods. %Thus, our approach provides a ``lazy'' alternative to the dominating Bayesian inference approaches for the unknown or hard-to-model scenarios.

%So far, our approach is limited for single target tracking% and has not explicitly taken into account the practical issues such as multiple-target interacting
\SPrevis{The future work can be threefold: The first is to extend the T-FoT approach} to the scenario of an unknown number of targets, i.e., find multiple T-FoTs that fit best the measurement sequences over time. Such an extension is nontrivial as measurement-to-track association is involved, which is challenging whenever target tracks are interacting with or approaching each other. % \citep{Vo15mtt}.
\SPrevis{The second is to extend the T-FoT approach for extended target or even swarm target tracking by utilizing extension/swarm feature estimation approaches.
The third is to consider a decentralized/large-scale sensor network for which an interesting issue would be continuous-time trajectory fusion \citep{Li22chapter}. All of these aim at a systematic data-driven and self-contained approach to multi-sensor multitarget detection and tracking.} % for consensus or synchronization. %, in addition to false and missing data. 8

%\begin{center}
%  \revis{Appendixes are new}
%\end{center}

\appendix
\section{Bound on Uniformly-distributed Clutter Rate}
\label{appendix_Clutter}
Disregarding the boundary issue of the surveillance region, %and consider the clutter that is uniformly distributed over the whole region. % and treat all local areas in the surveillance region the same.
the probability for a clutter point falling within a distance $d_o$ to the target in the region (which corresponds to a circle around the target) is simply given by
\begin{equation}\label{eq:circleProb}
  p_1 = S^{-1} \pi d_o^2,
\end{equation}
where \newrev{$S$ is the area of the entire surveillance region and} $\pi d_o^2$ gives the area of the circle with radius $d_o$.

Now, consider $r_c$ clutter points that are independently, uniformly generated over the region. The probability for all clutter points falling further than $d_o$ to the target is given by
\begin{equation}\label{eq:circleProb_rc}
  p_r = (1-p_1)^{r_c} = (1-S^{-1} \pi d_o^2)^{r_c}.
\end{equation}

To ensure no clutter generated within a distance $d_o$ to the target in order to avoid a FA within clustering under the confident probability $p_r$, the clutter rate needs to satisfy
\begin{equation}\label{eq:bound_rc}
 r_c \leq \log_{1-S^{-1} \pi d_o^2}(p_r).
\end{equation}

For instance, if $p_r=0.95, p_1 = 0.01$ (corresponding to an area that is 1\% of the whole area), then $\log_{0.99}(0.95) \approx 5.10$. That is, we have 95\% confidence that uniformly-distributed clutter points should not lie in any region whose area is 1\% of the whole area, when the clutter rate $r_c<5.1$. %In our approach, we set $d_o$ by the average distance between the measurements of the target (if detected) between two successive sensing scans. By this, no clutter is closer to the target than the real measurements of the target between two successive sensing scans at a confidence level $p_r$.

\section{Probabilistic Distance between IID Variables}
\label{appendix_Dist_Gauss_Variable}
For $D$-dimensional Gaussian-random variables $\mathbf{a}$ and $\mathbf{b}$ which are independently identically distributed with covariance $\mathbf{R}$. The probability that they have distance within $\tau$ times the standard deviation of their distribution is given by \citep{Ye00}
\begin{equation} \label{eq:confidence_Normal}
\mathrm{Pr}\big[(\mathbf{a}-\mathbf{b})^\mathrm{T} \mathbf{R}^{-1}(\mathbf{a}-\mathbf{b}) \leq \tau^2 \big] \leq \gamma \Big(\frac{D}{2},\frac{\tau^2}{2}\Big),
\end{equation}
where $\gamma(a,b):=\int_0^bt^{a-1}e^{-t}dt$ is the lower incomplete Gamma function.% and $d_\mathbf{x}$ is the dimension of the state.

When the variables are distribution free (i.e., unknown and probably
non-Gaussian) then the probability that the variable lies within the standard deviation of the state
estimate can still be bounded using, for example,
the Chebyshev inequality as follows
\begin{equation} \label{eq:Chebyshev}
\mathrm{Pr}\big[(\mathbf{c})^\mathrm{T} (\mathbf{c}) \geq \epsilon \big] \leq \frac{\Sigma_{\mathbf{c}}}{\epsilon^2},
\end{equation}
where $\mathbf{c}$ is a zero-mean random variable with variance $\Sigma_{\mathbf{c}}$ and $\epsilon >0$.

A tighter bound is given by the Chebyshev-type inequality,
Vysochanski\"{\i}-Petunin inequality \citep{Vysochanskii80}, when it is
known that the distribution is unimodal, i.e.,
\begin{equation} \label{eq:confidence_free}
\mathrm{Pr}\big[(\mathbf{a}-\mathbf{b})^\mathrm{T} \mathbf{R}^{-1}(\mathbf{a}-\mathbf{b}) \leq \tau^2 \big] \geq 1-\frac{4D}{9\tau^2}.
\end{equation}
%The choice of $\tau$ depends on its indication that the successive two detections of the target should not be more than $\tau_1$ times the measurement noise standard derivation if the target does not move. When the measurement noise is additive, zero-mean Gaussian,

\section*{Acknowledgment}
This work was supported in part
by the National Natural Science Foundation of China under Grant 62071389, by the Key Laboratory Foundation of
National Defence Technology under Grant JKWATR-210504, by the Natural Science Basic Research Program of Shaanxi and by the JWKJW Foundation .
The authors would like to thank Prof. P. Djuri\'c and Prof. F. Hlawatsch for their insightful discussion on the work.

\bibliographystyle{model1-num-names} %{IEEEtran}
\bibliography{TTinDark}

\end{document}